\documentclass[doublespacing]{elsart}

\usepackage{epsfig}

\newcommand{\be}{\begin{equation}}
\newcommand{\ee}{\end{equation}} \def\la{\mathrel{\mathpalette\fun
<}} 
\def\fun#1#2{\lower3.6pt\vbox{\baselineskip0pt\lineskip.9pt
\ialign{$\mathsurround=0pt#1\hfil
##\hfil$\crcr#2\crcr\sim\crcr}}}

\newcommand{\ver}{\mbox{\boldmath${\rm r}$}}

\newcommand{\vep}{\mbox{\boldmath${\rm p}$}}

 \newcommand{\lan}{\langle}
 \newcommand{\ran}{\rangle}
\newcommand{\lll}{\langle}
 \newcommand{\rrr}{\rangle}

\begin{document}
\begin{frontmatter}



\title{The Feynman-Schwinger representation in QCD}

\author[a,b]{Yu. A. Simonov}
\address[a]{ Research Center, ITEP, Moscow, Russia}
\address[b]{Jefferson Laboratory,Newport News,VA 23606,USA}

and

\author[c,d,e]{J. A. Tjon}
\address[c]{ITP, University of Utrecht, 3584 CC Utrecht, The Netherlands}
\address[d]{KVI, University of Groningen, 9747 AA Groningen, The Netherlands}
\address[e]{Department of Physics, University of Maryland, College
Park, MD  20742, USA}

\date{\today}

\begin{abstract}
The proper time path integral representation is derived explicitly
for Green's functions in QCD. After an introductory analysis
of perturbative properties, the total gluonic field is separated
in a rigorous way into a nonperturbative background and valence 
gluon part. For nonperturbative contributions the background
perturbation theory is used systematically, yielding two types of
expansions,illustrated by direct physical applications. 
As an application, we discuss the collinear singularities in the 
Feynman-Schwinger representation formalism.
Moreover, the generalization to nonzero temperature is made and 
expressions for partition functions in perturbation theory and 
nonperturbative background are explicitly written down.

\end{abstract}


\end{frontmatter}

\section{Introduction}

The present stage of development of field theory in general and of
QCD in particular requires the exploitation of nonperturbative
methods in addition to summing up perturbative series. This calls
for specific methods where the dependence on vacuum fields can be made
simple and explicit. A good example is provided by the so-called
Fock-Feynman-Schwinger representation (FSR) based on the
Fock-Schwinger proper time  and Feynman path integral
formalism\cite{1,2}. For QED asymptotic estimates the FSR was
exploited in Ref. \cite{3}.
Later on this formalism was used in Ref. \cite{4} for QCD and rederived
in the framework of the stochastic background method in Ref. \cite{5}.
(For a review see also Ref. \cite{6}.)

More recently some modification of the FSR was suggested in Ref. \cite{7}.
The one-loop perturbative amplitudes are especially convenient for the FSR. 
These amplitudes were extensively studied in Ref. \cite{8} and a convenient 
connection to the string formalism was found in Ref. \cite {9}.
Important practical applications especially for the effective action in QED
and QCD are contained in Ref. \cite{10}.
Moreover, the first extension of FSR to nonzero temperature field
theory was done in Refs. \cite{11,12}. This forms
the basis of a systematic study of the role of
nonperturbative (NP) configurations in the temperature phase
transitions\cite{11,13}.

One of the most important advantages of the FSR is that it allows to
reduce  physical amplitudes to weighted integrals of 
averaged Wilson loops. Thus the fields (both perturbative and NP)
enter only through Wilson loops. For the latter case one can apply
the cluster expansion method\cite{14}, which allows to sum up a
series of approximations directly in the exponent.
As a result we can avoid the summation of Feynman diagrams to get
the asymptotics of form factors\cite{15}.
The role of FSR in the treatment of NP effects is more crucial. In
this case one  can develop a powerful method of background
perturbation theory\cite{16} treating the NP fields as a 
background\cite{17}.

In the present paper some of these problems will be treated
systematically and in full detail, yielding a overall picture of
the role of FSR in QCD. We will in particular focus on the
relationship between the standard perturbative expansion and the
FSR based expansion which clarify the important role of nonperturbative
configurations in the vacuum.
The previous publication of the authors on FSR in Ref. \cite{18} was
devoted to  QED and $\varphi^3,\varphi^4$ theories, and many basic
formulas of FSR are already contained there. 
The later development of FSR in the framework of the field theory 
can be found in Ref. \cite{19}.
It has in particular been used successfully to reconstruct exact 
solutions of the one and two-particle Greens' function for $\varphi^3$
theory and scalar QED in the quenched approximation\cite{20,21,22,23,24,25}.
A review of the FSR applications to perturbation theory 
in QCD and a discussion of the connection between FSR and world-line
formalism of Refs. \cite{8,9,10} can be found in Ref. \cite {26}.

In the next section we describe how to derive the FSR formalism for the
case of QCD. In section 3 we discuss the relationship between the 
usual perturbative expansion and the FSR. 
Section 4 deals with the study of two ways
to determine the Green's function, depending on the physical situation.
One consists of an expansion in the perturbative fields and the 
other one is treating the nonperturbative fields as a correction.
As applications we address in sections 5 and 6 the problems of collinear
singularities and the finite temperature field theory in the FSR 
formalism, while some concluding remarks are made in the last section.

\section{General form of FSR in QCD}

Let us consider a scalar particle $\varphi$ (e.g. a Higgs boson) interacting
with a nonabelian vector potential $A$, where the Euclidean Lagrangian
is given by
\be
L_\varphi = \frac12 |D_\mu \varphi|^2 + \frac12 m^2 |\varphi|^2
\equiv \frac12 |(\partial_\mu-ig
A_\mu) \varphi |^2 + \frac12 m^2 |\varphi|^2,
\label{1}
\ee
Using the Fock--Schwinger proper time representation 
the two-point Green's function of $\varphi$ can be written in the
quenched approximation as
\be
G(x,y) = (m^2-D^2_\mu)^{-1}_{xy} = \langle x | P \int^\infty_0 ds
e^{-s(m^2-D^2_\mu)}|y\rangle.
\label{2}
\ee
To obtain the FSR for $G$ a second step is needed.
As in Ref.~\cite{1} the matrix element in Eq.~(\ref{2})
can be rewritten in the form of a path integral, treating $s$ as the
ordering parameter. Note the difference of the  integral (\ref{2}) from 
the case of the Abelian QED treated in Refs.~\cite{1,3,8}:
$A_\mu$ in our case is a matrix operator $A_\mu(x)=
A_\mu^a(x)T^a$. It does not commute for different $x$. Hence
the ordering operator $P$ in Eq.~(\ref{2}). The precise meaning of $P$
becomes more clear in the final form of a path integral
\be
G(x,y) =\int^\infty_0\!\! ds (Dz)_{xy}\, e^{-K} P \exp\left(ig \!\!\int^x_y
\!\!A_\mu (z) dz_\mu\right),
\label{3}
\ee
where $K= m^2s+ \frac14 \int^s_0 d\tau (\frac{dz_\mu}{d\tau} )^2$.
In Eq.~(\ref{3}) the functional integration measure can be written as
\be 
(Dz)_{xy} \simeq \lim_{N\to \infty} \prod^N_{n=1}\!\int\!\!
\frac{d^4 z(n)}{(4\pi\varepsilon)^2}\! \int\!\! \frac{d^4p}{(2\pi)^4}
e^{ip \left(\sum^N_{n=1}\! z(n)-(x-y)\right)}
\label{4}
\ee
with $N\varepsilon =s$. 
The last integral in Eq.~(\ref{4}) ensures that the path $z_\mu(\tau),
0\leq \tau\leq s$, starts at $z_\mu(0) = y_\mu$ and ends at
$z_\mu(s) =x_\mu$. The form of Eq.~(\ref{3}) is the same as in the case
of QED except for the ordering operator $P$ which provides a
precise meaning to the integral of the noncommuting matrices
$A_{\mu_1}(z_1), A_{\mu_2}(z_2)$ etc. In the case of QCD the forms
(\ref{3}) and (\ref{4}) were introduced in Refs. \cite{4,5}.

The FSR, corresponding to a description in terms of particle
dynamics is equivalent to field theory, when all the
vacuum polarisation contributions are also included\cite{4,5,10},
i.e.
\begin{eqnarray}
&&\sum_{N=0}^{\infty} \frac{1}{N!} \prod_{i=1}^{N} \int\! \frac{ds_i}{s_i}
\!\int\! (Dz_i)_{xx} \exp(-K)\; P\exp\left(ig\! \!\int\limits_y^x
\!\!A_{\mu}(z) dz_{\mu}\right) 
\nonumber\\[1mm]
&&= \int\! D\varphi \exp\left(\!-\!\!\int\! \! d^4 x L_{\varphi}(x)\right ).
\label{bss}
\end{eqnarray}
Both sides are equal to vacuum-vacuum transition amplitude in the presence
of the external nonabelian vector field and hence to each other.
For practical calculations proper regularization of the above equation has
to be done.
The field $A_{\mu}$ in  Eq.~(\ref{1}) can be considered
as a classical external field or as a quantum one.
In the latter case the Green's functions $\lll A .. A \rrr$
induce nonlocal current-current interaction terms in the l.h.s. of
Eq.~(\ref{bss}).
Such terms can also be generated by the presence of a $\varphi$-field potential,
$V(|\varphi|)$ in the r.h.s. of   Eq.~(\ref{bss}). 

The advantage of the FSR in this case follows from the very
clear space-time picture of the corresponding dynamics in terms of
particle trajectories.
This is especially important if the currents can be treated
as classical or static (for example, in the heavy quark case).
   The mentioned remark on usefulness of the FSR (\ref{3}) becomes 
clear when one
   considers the physical amplitude, e.g. the Green's function of the
   white state $tr(\varphi^+(x) \varphi(x))$ or its nonlocal version
   $tr[\varphi^+ (x) \Phi(x,y) \varphi(y)]$, where $\Phi(x,y)$ -- to be
   widely used in what follows -- is the parallel transporter along some
   arbitrary contour $C(x,y)$
   \be
   \Phi(x,y) =P\exp\left(ig\!\! \int^x_y \!\!A_\mu(z) dz_\mu\right).
   \label{5}
   \ee
   One has by standard rules
\begin{eqnarray}
 &&G_\varphi (x,y) =\left\langle
    tr\left[\varphi^+(x) \varphi(x)\right]\,
    tr \left[\varphi^+(y)
\varphi(y)\right]\right\rangle_A\nonumber\\[1mm]
 &&
     =\int^\infty_0\!\!ds_1\! \int^\infty_0\!\! ds_2
   (Dz)_{xy}(Dz')_{xy}\, e^{-K-K'} \left\langle W\right\rangle_A + \ldots
   \label{6}
\end{eqnarray}
   where dots stand for the disconnected part,
   $\langle G_\varphi (x,x) G_\varphi (y,y)\rangle_A$. We 
have used the fact that the propagator for the
   charge-conjugated field $\varphi^+$ is proportional to
$\Phi^{\dagger}(x,y) = \Phi(y,x)$.
%\be
%   \Phi_C(x,y) =P\exp (ig \int^x_y A_\mu^{(C)} (z) dz_\mu), ~~
%   A_\mu^{(C)}=-A_\mu^{(T)}
%   \label{7}
%   \ee
   Therefore the ordering $P$ must be inverted, $\Phi^{\dagger}(x,y)
   =P \exp (ig \int^y_x A_\mu (z) dz_\mu)$.
Thus all dependence on $ A_\mu$ in $G_\varphi$ is reduced to the
Wilson loop average
\be
\left\langle W\right\rangle_A=\left\langle tr \,P_C \exp ig \!\!\int_C
 \!\!  A_\mu (z) dz_\mu\right\rangle_A.
     \label{8}
     \ee
      Here $P_C$ is the ordering  around the closed loop
      $C$ passing through the points x and y, the loop
      being made of the paths $z_\mu(\tau)$,
      $z'_\mu(\tau')$ and to be integrated over.

The FSR can also be used to describe the quark and gluon propagation.
Similar to the QED case, 
      the fermion (quark) Green's function in the presence of an
      Euclidean external gluonic field can be written as
\begin{eqnarray}
     \label{9}
 &&     G_q(x,y)= \langle \psi(x) \bar \psi(y)\rangle_q=\langle x
      |(m+\hat D)^{-1}|y\rangle 
\nonumber
\\[1mm]
&&     = \langle x|(m-\hat D)(m^2-\hat D^2)^{-1}|y\rangle
\nonumber
\\
&&  =  (m-\hat D)\!\int^\infty_0\!\! ds (Dz)_{xy} e^{-K} \Phi_\sigma(x,y)\,,
\end{eqnarray}
where $\Phi_\sigma$ is the same as was introduced in Ref.~\cite{1} except for
the ordering operators $P_A,~P_F$
      \be
      \Phi_\sigma (x,y) = P_A\exp\left(ig \int^x_y A_\mu
      dz_\mu\right)\,P_F 
      \exp \left(g\int^s_0 d\tau \sigma_{\mu\nu} F_{\mu\nu}\right)
      \label{10}
      \ee
with $F_{\mu\nu}=\partial_{\mu} A_{\nu}-\partial_\nu A_{\mu}-ig [A_\mu,A_\nu]$ 
and $\sigma_{\mu\nu}=\frac{1}{4i}
      (\gamma_\mu\gamma_\nu-\gamma_\nu\gamma_\mu)$, while $K$ and
      ($Dz)_{xy}$ are defined in Eqs.~(\ref{3}) and (\ref{4}). Note that
      operators $P_A, P_F$ in Eq.~(10) preserve the proper ordering of
      matrices $A_\mu$ and $\sigma_{\mu\nu}  F_{\mu\nu}$
      respectively. Explicit examples are considered below.

      Finally we turn to the case of FSR for the valence gluon
      propagating in the background nonabelian field. 
Here we only quote the result for the gluon Green's function in the background
      Feynman gauge\cite{5,17}. We have
      \be
      G_{\mu\nu} (x,y) =
      \langle x|(D^2_\lambda \delta_{\mu\nu}-
      2ig F_{\mu\nu})^{-1}
      |y\rangle
      \label{11}
      \ee
      Proceeding in the same way as for quarks, we obtain the FSR
      for the gluon Green's function
      \be
      G_{\mu\nu} (x,y) =
      \int^{\infty}_0 ds (Dz)_{xy} e^{-K_0}\Phi_{\mu\nu}(x,y),
      \label{12}
      \ee
      where we have defined
\begin{eqnarray}
 K_0&\!\!=&\!\!\frac14 \int^\infty_0 \left
(\frac{dz_\mu}{d\tau}\right)^2
      d\tau, \nonumber\\[1mm]
\Phi_{\mu\nu} (x,y) &\!\!\!\!=&\!\!\!\!\left [P_A\exp
\left(ig\!\!\int^x_y\!\! A_\lambda
      dz_\lambda\right)P_F
\exp \left(2g\!\!\int^s_0\!\! d\tau F_{\sigma\rho}
      (z(\tau))\right)\right]_{\mu\nu}\! .
      \label{13}
\end{eqnarray}
      Now in the same way as is done above for scalars in
Eq.~(\ref{6}), we may consider a Green's function,
corresponding to the physical
      transition amplitude from a white state of $q_1, \bar q_2$ to
      another white state consisting of $q_3,\bar q_4$. It is given by
      \be
      G^\Gamma_{q\bar q} (x,y) =
      \langle G_{q} (x,y)
       \Gamma G_{\bar q} (x,y)\Gamma-
       G_{ q} (x,x)
              \Gamma G_{\bar q} (y,y)\Gamma\rangle_A,
              \label{14}
              \ee
where $\Gamma$ describes the vertex part for the interaction between 
the $q,\bar{q}$ pair in the
meson. The first term on the r.h.s. of Eq.~(\ref{14}) can be
              reduced to the same form as in Eq.~(\ref{6}) but with the
              Wilson loop containing ordered insertions of the
              operators $\sigma_{\mu\nu} F_{\mu\nu}$ (cf. Eq.~(\ref{10})).

\section{Perturbation theory in the framework  of FSR. Identities and partial summation}

In this section we  discuss in detail how the usual results of
perturbation theory follow from FSR. It  is useful to establish such
a general  connection between the perturbation series (Feynman
diagram technique) and FSR.
At the same time the FSR
presents a unique possibility to sum up Feynman diagrams in a very
simple way, where the final result of the summation is written in an
exponentiated way\cite{15,17}. This method will be discussed  in
the next section.

Consider the FSR for the  quark Green's function. According to
Eq. (\ref{9}), the 2-nd order of  perturbative expansion of
Eq. (\ref{10}) can be written as
%\begin{eqnarray}
$$G_q(x,y)= (m-\hat D) \!\int^\infty_0\!\!  ds 
\!\int^\infty_0 \!\!  d\tau_1\!
\int^{\infty}_0\!\!  d\tau_2\, e^{-K} (Dz)_{xu} d^4u (Dz)_{uv}d^4v(Dz)_{vy}
$$
\be
\times
 \left(ig A_\mu(u)\dot u_\mu+g\sigma_{\mu\nu} F_{\mu\nu} (u)\right)
\left(ig A_\nu(v)\dot v_\nu+ g\sigma_{\lambda\sigma}
F_{\lambda\sigma} (v)\right),
\label{15}
%\end{eqnarray}
\ee
 where we have used the identities
%\begin{eqnarray}
\be
%&&
(Dz)_{xy} = (Dz)_{xu(\tau_1)} d^4u(\tau_1)(Dz)_{u(\tau_1)v(\tau_2)} 
d^4 v(\tau_2) (Dz)_{v(\tau_2)y},
 \label{16}\\[1mm]
\ee
%&&
\be 
\int^\infty_0\!\!ds\!\!\int^s_0\!d\tau_1\!\!\int^{\tau_1}_0\!\!d\tau_2
 f(s,\tau_1,\tau_2)=
\int^\infty_0\!ds\!\int^\infty_0 d\tau_1\!\int^\infty_0\!\!d\tau_2
 f(s+\tau_1+\tau_2,\tau_1+\tau_2,\tau_2).
 \label{17}
%\end{eqnarray}
\ee
We can also expand only in the color magnetic moment interaction ($\sigma
F$). This is useful when the spin-dependent interaction can be
treated perturbatively, as it is in most cases for mesons and
baryons (exclusions are Goldstone bosons and nucleons,
where the spin interaction is very important and interconnected with
chiral dynamics). In this case we obtain
to the second order in ($\sigma F$)
%\begin{eqnarray}
$$
G_q^{(2)}(x,y) = i(m-\hat
D)\!\!\int^\infty_0\!\!ds\!\!\int^\infty_0\!\! d\tau_1\!\!\int^\infty_0
\!\!d\tau_2\, e^{-m^2_q(s+\tau_1+\tau_2)-K_0-K_1-K_2} $$
\be
(Dz)_{xu}\Phi (x,u) g(\sigma F(u)) d^4 u(Dz)_{uv}
\Phi(u,v) g(\sigma F(v)) d^4v(Dz)_{vy}. \label{18} 
%\end{eqnarray}
\ee
In another way it can be written as
\begin{eqnarray}
G^{(2)}_q(x,y) =&&i(m-\hat
D) (m^2_q-D^2_\mu)^{-1}_{xu} d^4u~g(\sigma
F(u))(m^2_q-D^2_\mu)^{-1}_{uv} d^4v
\nonumber
\\
&&\times g(\sigma F(v)) (m^2_q-D^2_\mu)^{-1}_{vy}. 
\label{19}
\end{eqnarray}
 Here $( m^2_q-D^2_\mu)^{-1}$ is the Green's function of a scalar
 quark in the external gluonic field $A_\mu$.
 This type of expansion is useful also for the study of small-$x$
 behavior of static potential, since
 the correlator $\lan\sigma F(u) \sigma F(v)\ran$ plays an
 important role there.

 However, in establishing the general connection between perturbative
 expansion for Green's functions in FSR and expansions of
 exponential $\Phi_\sigma$ in Eq.~(\ref{10}), one encounters a
 technical difficulty since the coupling constant $g$ enters in three
 different ways in FSR:
%\begin{enumerate}
%\item 

1. in the factor $(m-\hat D)$ in front of the integral in Eq.~(\ref{9})

%\item 
2. in the parallel transporter (the first exponential in Eq.~(\ref{10}))
%\item

3. in the  exponential of $g(\sigma F)$.
%\end{enumerate}

\noindent
Therefore it is useful to compare the two expansions in the operator form
\begin{eqnarray}
  (m+\hat D)^{-1}=&& (m+\hat \partial - ig \hat A)^{-1}=(m+\hat
  \partial)^{-1}+ (m+\hat \partial)^{-1} ig \hat A (m+\hat
  \partial)^{-1}
\nonumber
\\
  &&+ (m+\hat \partial)^{-1} ig \hat A(m+\hat \partial)^{-1} ig \hat
  A(m+\hat \partial )^{-1}+...
  \label{20}
\end{eqnarray}
  and the FSR 
  \be
  (m+\hat D)^{-1} = (m-\hat D) (m^2-\partial^2)^{-1}
  \sum^\infty_{n=o}(\delta(m^2-\partial^2)^{-1})^n,
  \label{21}
  \ee
  where we have introduced 
  \be
  \delta=-ig (\hat A\hat \partial+\hat \partial \hat A) - g^2 \hat
  A^2\equiv \hat D^2-\partial^2.
  \label{22}
  \ee
To see how the expansion (\ref{21}) works, using 
$\hat D=\hat \partial -ig \hat A$,
Eq.~(\ref{21}) becomes
$$
  (m+\hat D)^{-1}=[(m+\hat \partial)^{-1} +ig \hat A
  (m^2-\partial^2)^{-1}]\sum^\infty_{n=0}
  [\delta(m^2-\partial^2)^{-1}]^n
  $$
Separating out the first term we may rewrite this as
\begin{eqnarray}
 (m+\hat D)^{-1}=&&(m+\hat \partial)^{-1}
\nonumber
\\
&& + (m+\hat \partial)^{-1} ig \hat
  A(m-\hat D) (m^2-\partial^2)^{-1}\sum^\infty_{n=0}
     [\delta(m^2-\partial^2)^{-1}]^n
\label{23}
\end{eqnarray}

%  \be
%  =(m+\hat \partial)^{-1}+ (m+\hat\partial)^{-1} ig \hat A(m+\hat D)^{-1}.
%     \label{23}
%     \ee
The last three factors in Eq.~(\ref{23}) are the
same as occurring in Eq.~(\ref{21}). As a consequence the formal iteration of 
the resulting equation for the Greens' function
reproduces the same series as in Eq.~(\ref{20}), showing the equivalence of the
two expansions.

It is important to note that each term in the expansion in powers of
$\delta$, after
 transforming the operator form of Eq.~(\ref{21}) into the integral
 form of FSR, becomes an expansion of the  exponential
 $\Phi_\sigma$ in Eq.~(\ref{10}) in powers of $g$.
 The second order term
 of this expansion was written down before in Eq.~(\ref{15}).

It is our purpose now to establish the connection
between the expansion (\ref{21}), (\ref{23}) and the
expansion of the exponential $\Phi_\sigma$ in 
Eq.~(\ref{10}) in the quark propagator (\ref{9}).
We can start with term linear in $\hat A$ and write
(for the Abelian case see Appendix B of Ref. \cite{18})
 \be 
G_q^{(1)} = ig \int G^{(0)}_q(x,z(\tau_1)) d^4
z\frac{\xi_\mu(n)}{\varepsilon}
 \bar A_\mu(\tau_1) G_q^{(0)}(z(\tau_1), y),
\label{24}
\ee
where the notation is clear from the general representation of $G_q$,
given by Eq.~(\ref{9})
\be
G_q(x,y)
 =\int^\infty_0 ds e^{-sm^2_q} \prod^N_{n=1}
\frac{d^4\xi(n)}{(4\pi\varepsilon)^2}
\exp\left[-\sum^N_{n=1} \frac{\xi^2(n)}{4\varepsilon} \right ]
\Phi_\sigma (\bar A,\xi) 
\label{25} 
\ee 
with 
$\xi (n) = z(n) - z(n-1),~~ \bar A_\mu (n) =\frac12[A_\mu(z(n))+A_\mu(z(n-1))]$ 
and 
\be 
\Phi_\sigma (\bar A, \xi) =P\exp \{ ig \sum^N_{n=1} \bar
A_\mu(n) \xi_\mu(n) + g\sum^n_{n=1} \sigma_{\mu\nu}
F_{\mu\nu}(z(n))\varepsilon \}.
\label{26} 
\ee 
Representing $\xi(n)$ in Eq.~(\ref{24}) as $\frac12(\xi_\mu(L)+\xi_\mu(R))$,
where
 $\xi_\mu(L)$ refers to the integral over $\xi_\mu$ in
 $G_q^{(0)}$
to the left of $\xi_\mu$ in Eq.~(\ref{24}) and $\xi_\mu(R)$ to the
integral in $G_q^{(0)}$ standing to the right of $\xi_\mu$, we
obtain 
\be 
\int\xi_\mu(n)\frac{d^4\xi(n)}{(4\pi\varepsilon)^2}
e^{ip\xi-\frac{\xi^2}{4\varepsilon}}=-i\frac{\partial}{\partial
p_\mu}  e^{-ip^2\varepsilon}=2 ip_\mu\varepsilon
e^{-p^2\varepsilon}. \label{27}
\ee
Thus Eq.~(\ref{24}) in momentum space becomes 
\be
G_q^{(1)}=-gG^{(0)}_q(q)\lan q|p_\mu A_\mu+A_\mu p_\mu|q'\ran
G^{(0)}_q(q') 
\label{28} 
\ee
In a similar way the second order term from the  coinciding
arguments yields
\be 
G^{(2)}_q(coinc) = -g^2\int G^{(0)}_q(x,z) A^2_\mu(z) d^4 z
G_q^{(0)}(z,y). 
\label{29} 
\ee 
Finally, the first order expansion
of the term $\sigma_{\mu\nu} F_{\mu\nu}$ in Eq.~(\ref{10}) yields the
remaining missing component of the combination $\delta$, Eq.~(\ref{22}), 
which can be  rewritten as 
\be 
\delta=-ig (A_\mu\partial_\mu+\partial_\mu
A_\mu)-g^2 A^2_\mu+g\sigma_{\mu\nu} F_{\mu\nu}. 
\label{30} 
\ee
Hence the second term in the expansion (\ref{21}) 
\begin{eqnarray} 
(m+\hat D)^{-1} =&& (m-\hat D) (m^2-\partial^2)^{-1}
\nonumber
\\
&&+(m-\hat
D)(m^2-\partial^2)^{-1}\delta(m^2-\partial^2)^{-1}+... 
\label{31}
\end{eqnarray}
is exactly reproduced by the expansion of the FSR (\ref{9}),
where in the first exponential $ \Phi_\sigma$ in Eq.~(\ref{9}) one
keeps terms of the first and second order, $O(gA_\mu)$ and
$O((gA_\mu)^2)$, while in the second exponential one keeps only
the first order term $O(g\sigma_{\mu\nu}F_{\mu\nu})$. It is easy
to see that this rule can be generalized to higher orders of
the expansion in $\delta$ in Eq.~(\ref{21}) as well.

 \section{Perturbative $vs$ nonperturbative: two types of
 expansion }

 As was discussed in Section 2, gluons can also be
  considered in FSR. To make this statement explicit and to prove Eqs. (\ref{11}-\ref{13})
 written for the gluon Green's functions, we can use the background perturbation theory\cite{16}.
  As in Ref. \cite{17}  we combine the perturbative field
 $a_{\mu}$  and NP degrees of freedom $B_{\mu}$ in one gluonic field
 $A_{\mu}$, namely
 \be
 A_{\mu}=B_{\mu}+a_{\mu}
 \label{4.1}
 \ee
 Under gauge transformations  $A_{\mu}$ transforms as
 \be
 A_{\mu}\to A'_{\mu}=U^+(A_{\mu}(x)+\frac{i}{g}\partial_{\mu})U
\label{4.2}
 \ee
 At this point we must distinguish two opposite physical situations,
  which require different types of expansions. Consider
 first systems of small size, e.g.  heavy  quarkonia,
 which are  mostly governed by the color Coulomb interaction
 and have a radius of the order $(m\alpha_s(m))^{-1}$, where $m$ is
 the quark mass. For the ground state bottomonium this
 radius is around 0.2 fm and for charmonium 0.4 fm.

 In this case we  have the {\it first type of
 expansion:} at the zeroth order all gluon exhanges are
 taken into account (In practice the Coulomb contribution and the few
 first radiative  corrections), while in first order one
 treats the nonperturbative contribution as a correction.
 This  expansion is considered in detail  at the end of this
 Section.

 The {\it second type of expansion} takes into account
 NP interaction fully through NP vacuum correlators already
 in the zeroth order --this is the NP  background and in the
 next orders the usual background perturbation theory\cite{16,17}
is developed with necessary modifications.

\subsection{ Expansion in perturbative fields}

We start  this Section  with the second type of expansion
with some modifications due to the independent
 integral over the background field, as in the 't Hooft's
 identity\cite{17}.
It is convenient to impose on $a_{\mu}$ the background gauge 
condition\cite{16}
 \be
 D_{\mu}a_{\mu}=\partial_{\mu}a_{\mu}^a+g
 f^{abc}B^b_{\mu}a^c_{\mu}=0.
 \label{4.5}
 \ee
 In this case a ghost field has to be introduced. Defining 
$  D_{\lambda}^{ca} =
 \partial_{\lambda}\cdot \delta_{ca}+ g~f^{cba} B^b_{\lambda}
\equiv \hat{D}_{\lambda}$,
We can write the resulting partition function as
 \be
 Z=\frac{1}{N'}\int DB e^{\int J_{\mu}B_{\mu}d^4x}
 Z(J,B),
 \label{4.6}
 \ee
 where
\begin{equation}
Z(J,B) = \int Da ~det(\frac{\delta G^a}{\delta \omega^b}) \exp  \int d^4 x
[L_0 + L(a)-\frac{1}{2\xi}(G^a)^2 + J^a_{\mu} a^a_{\mu}].
\label{4.7}
\end{equation}
In Eq. (\ref{4.7}) we have 
\begin{equation}
L_0 = -\frac{1}{4}
 (F^a_{\mu \nu}(B))^2 
\label{4.8}
\end{equation}
and 
\be
L(a) = L_1(a)+L_2(a)+L_{int}(a) 
\label{la}
\ee
with
\begin{eqnarray}
L_1(a)&=&a^c_{\nu} D_{\mu}^{ca}(B) F^a_{\mu\nu}
\nonumber \\
L_2(a)&=&+\frac{1}{2} a_{\nu}(\hat{D}^2_{\lambda}\delta_{\mu\nu} -
\hat{D}_{\mu}\hat{D}_{\nu} + ig \hat{F}_{\mu\nu}) a_{\mu}= \nonumber \\
&=&\frac{1}{2} a^c_{\nu}[D_{\lambda}^{ca}D_{\lambda}^{ad}
\delta_{\mu\nu} - D_{\mu}^{ca}D_{\nu}^{ad} -
g~f^{cad}F^a_{\mu\nu}]a^d_{\mu}~~,
\nonumber
\\
L_{int} &=& -\frac{1}{2} g~(D_{\mu}(B)a_{\nu} -D_{\nu}(B)a_{\mu})^a
f^{abc} a_{\mu}^b a_{\nu}^c - \frac{1}{4} g^2 f^{abc} a_{\mu}^b
a_{\nu}^c f^{aef} a^e_{\mu} a^f_{\nu}.
\label{4.9}
\end{eqnarray}
$G^a$ in Eq. (\ref{4.7}) is the background gauge condition 
\begin{eqnarray}
G^a = \partial_{\mu}a^a_{\mu} + g f^{abc} B^b_{\mu} a^c_{\mu} =
(D_{\mu} a_{\mu})^a.
\label{4.10}
\end{eqnarray}
The ghost vertex is obtained from $\frac{\delta G^a}{\delta \omega^b}
= (D_{\mu}(B)D_{\mu} (B+a))^{ab}$~\cite{16} to be
\begin{eqnarray}
L_{ghost}=-\theta^+_a (D_{\mu}(B)D_{\mu} (B+a))^{ab} \theta_b,
\label{4.11}
\end{eqnarray}
$\theta$ being the ghost field.
The  linear part of the  Lagrangian $L_1$ disappears if 
$B_{\mu} $ satisfies the classical equations of motion. Here we  do not
impose this condition on $B_{\mu}$. However, it was  shown in
Ref. \cite{17} that $L_1$ gives no important contribution

We now can identify the propagator of $a_{\mu}$ from the quadratic terms
in the Lagrangian $L_2(a) - \frac{1}{2\xi}(G^a)^2$. We get
\begin{eqnarray}
G^{ab}_{\nu\mu} = [\hat{D}^2_{\lambda} \delta_{\mu \nu} -
\hat{D}_{\mu}\hat{D}_{\nu} + ig \hat{F}_{\mu\nu} +\frac{1}{\xi}
\hat{D}_{\nu} \hat{D}_{\mu}]^{-1}_{ab}.
\label{4.12}
\end{eqnarray}
It will be convenient sometimes to choose $\xi =1$ and end up with
the well-known form of the propagator in -- what one would call --
the background Feynman gauge
\begin{eqnarray}
G^{ab}_{\nu\mu} = (\hat{D}^2_{\lambda}\cdot  \delta_{\mu \nu} -
2ig \hat{F}_{\mu\nu}) ^{-1}
\label{4.13}
\end{eqnarray}
This  is exactly the form of gluon propagator used in Eq. (\ref{11}).
Integration over the ghost and gluon degrees of freedom in Eq. (\ref{4.7})
yields
  \begin{eqnarray}
  Z(J,B)=&&const (det W(B))^{-1/2}_{reg}[det
  (-D_{\mu}(B)D_{\mu}(B+a)]_{a=\frac{\delta}{\delta J}}
\nonumber
\\
&&  \times \{ 1+\sum^{\infty}_{l=1}
  \frac{S_{int}(a=\frac{\delta}{\delta J})^{l}}{l!}
  exp (-\frac{1}{2}JGJ)\left|_{J_{\nu}=D_{\mu}(B)F_{\mu\nu}(B)}
\right.,
  \label{4.14}
  \end{eqnarray}
where $S_{int}$ is the action corresponding to $L(a)$ and
$G$ is defined in Eq. (\ref{4.13}).

Let us mention the convenient gauge prescription
for gauge transformations of the fields $a_\mu,~B_\mu$.
Under the gauge transformations the fields transform as 
 \be
 a_{\mu}\to U^+a_{\mu}U
 \label{4.3},
 \ee
 \be
 B_{\mu}\to U^+(B_{\mu}+\frac{i}{g}\partial_{\mu})U.
\label{4.4} 
\ee
All the terms in Eq. (\ref{4.7}), 
including the gauge fixing one $\frac{1}{2}(G^a)^2$ are gauge 
invariant. That was actually one of the aims put forward by 
't Hooft in Ref. \cite{16}. It
  has important consequences: 

(i)  Any amplitude in the perturbative
  expansion in $ga_{\mu}$ of Eqs. (\ref{4.7}) and (\ref{4.14}) corresponding to a
  generalized Feynman diagram, is separately gauge invariant (for
  colorless initial and final states of course).

  (ii)  Due to gauge invariance of all terms, the renormalization is
  specifically simple in the background field formalism\cite{16}, since
  the counterterms enter only in gauge--invariant combinations, e.g.
  $F^2_{\mu\nu}$. The $Z$--factors $Z_g$  and $Z_B$ are
  connected: $Z_gZ_B^{1/2}=1$.

  As a consequence, the quantities like $gB_\mu$,
  $gF_{\mu\nu}(B)$ are renormalization-group (RG) invariant.
  Consequently all background field correlators are also RG
  invariant and they can be considered on the same footing as the
  external momenta  in the amplitudes. This leads to a new
  form of solutions of RG equations, where
  $\alpha_s=\alpha_s(M_B)$,  $M_B$ being the (difference) of the
  hybrid excitations, typically $M_B\approx 1 GeV$. As a result a
  new phenomenon appears, freezing or saturation of $\alpha_s$ at
  large Euclidean distances. For more discussion see Ref. \cite{17} and
  recent explicit extraction of the freezing $\alpha_s$  from the
  spectra of heavy quarkonia\cite{37}. In the rest of this
  subsection we demonstrate how the background perturbation series
  (\ref{4.14}) works for the  meson Green's function. To this end
  we use Eq. (\ref{14}) and consider the flavour nonsinglet case to
  disregard the second term in Eq. (\ref{14}).

 Let us start with the meson Green's function and use the FSR
 for  both quark and antiquark.

\begin{eqnarray} 
G_M(x,y)=\langle tr \Gamma^{(f)}(m-\hat D)&&\int^\infty_0 ds
\int^\infty_0 d\bar s e^{-K-\bar K}
\nonumber
\\
&&(Dz)_{xy}(D\bar
z)_{xy}\Gamma^{(i)}(\bar m-\hat{\bar D}) W_F\rangle 
\label{87} 
\end{eqnarray}
Here the barred symbols refer to the antiquark and 
\be 
W_F=P_A P_F
\exp(ig\oint dz_\mu A_\mu) \exp(g \int^s_0
d\tau\sigma^{(1)}_{\mu\nu}F_{\mu\nu}) \exp (-g\int^{\bar
s}_0\sigma^{(2)}_{\mu\nu}F_{\mu\nu} d\bar\tau). 
\label{88} 
\ee 
In Eq. (\ref{87}) integrations over proper times $s, \bar s$ and
$\tau, \bar\tau$ occur, which also play the role of an ordering parameter
along the trajectory, $z_\mu=z_\mu(\tau),~~\bar z_\mu=\bar
z_\mu(\bar \tau)$.

It is convenient to go over to the actual time $t\equiv z_4$ of
the quark (or antiquark), defining the new quantity $\mu(t)$,
which will play a very important role in what follows \be 2\mu(t) =
\frac{dt}{d\tau},~~ t\equiv z_4(\tau).
 \label{89} 
\ee
For each
quark (or antiquark and gluon) we can rewrite the path integral
(\ref{87}) as (see Refs. \cite{29,33} for details)
\be \int^\infty_0 ds (D^4z)_{xy}~~... = const
\int D\mu(t) (D^3z)_{xy}~~... \label{90}
  \ee where
$(D^3z)_{xy}$ has the same form as in Eq. (\ref{4}) but with all
4-vectors replaced by 3-vectors. The path integral $D\mu(t)$
is supplied with the proper integration measure, which is  derived
from the free motion Lagrangian.

In general $\mu(t)$ can be a strongly oscillating function of $t$
due to the Zitterbewegung. In what follows we shall use the
stationary point method for the evaluation of the integral over
$D\mu(t)$, with the extremal $\mu_0(t)$ playing the role of
an effective or constituent quark mass. We shall see that in all
cases, where spin terms can be considered as a small perturbation,
i.e. for the majority of mesons, $\mu_0$ is positive and rather large
even for vanishing quark current masses $m,\bar m$, and the role
of the Zitterbewegung is small (less than 10\% from the comparison to
the light-cone Hamiltonian eigenvalues, see Refs. \cite{29,30} for
details).

Now the kinetic terms can be rewritten using Eq. (\ref{89}) as 
\begin{eqnarray}
K+\bar K=\int^T_0 dt\{ \frac{m^2} {2\mu(t)}&& +\frac{\mu(t)}{2}
[(\dot z_i(t))^2+1]
\nonumber
\\
&&+ \frac{\bar m^2}{2\bar \mu(t)} +\frac{\bar
\mu(t)}{2} [(\dot{\bar z_i}(t))^2+1]\}, 
\label{91} 
\end{eqnarray}
where $T=x_4-y_4$. In the spin-dependent factors the corresponding
changes are 
\be 
\int^s_0 d\tau \sigma_{\mu\nu} F_{\mu\nu}=
\int^T_0 \frac {dt}{2\mu(t)}\sigma_{\mu\nu} F_{\mu\nu}(z(t)).
\label{92} 
\ee 
In what follows in this section  we may
systematically do a perturbation expansion of the spin terms. They
contribute to the total mass corrections of the order of 10-15\%
for lowest mass mesons, while they are much smaller for the 
high excited states. This
perturbative approach fails however for pions (and kaons) where
the chiral degrees of freedom should  be taken into account. In this
case another equation should be considered\cite{31,32}.  

Therefore as a starting approximation we may use the Green's functions
of mesons made of spinless quarks. This amounts to neglecting in
Eqs. (\ref{87},\ref{88}) the terms $(m-\hat D), (\bar m-\hat {\bar D})$
and $\sigma_{\mu\nu}F_{\mu\nu}$.  As a result, we have
\be
G^{(0)}_M(x,y)= const \int D\mu(t) D\bar \mu(t) (D^3z)_{xy}
(D^3\bar z)_{xy} e^{-K-\bar K}\langle W\rangle .  
\label{93} 
\ee
The Wilson loop in Eq. (\ref{93}) contains both perturbative and NP
fields. It can be expanded as 
\begin{eqnarray} 
W(B+a) =&& W(B)
\nonumber
\\
&& + \sum^\infty_{n=1} (ig)^nW^{(n)}(B;x(1)...x(n)) a_{\mu_1}
dx_{\mu1}(1)... d x_{\mu_n}(n).
 \label{93a} 
\end{eqnarray}
After averaging over $a_\mu$, $B_\mu$ we obtain, keeping the
lowest correction term: 
\be 
\lan W(B+a)\ran_{B,a} = \lan
W(B)\ran_B-g^2\lan W^{(2)}(B; x, y)\ran dx dy~~+~~... ,
\label{93b} 
\ee
where the second term in Eq. (\ref{93b}) can be written as 
\begin{eqnarray}
-g^2W^{(2)} dxdy=&&
 -g^2 \int \Phi^{\alpha\beta} (x,y,B)
\nonumber
\\
&&\times t_a^{\delta\alpha} t^{\beta\gamma}_b G^{ab}_{\mu\nu}(x,y,B)
\Phi^{\gamma\delta}(y,x,B) dx_{\mu}dy_\nu ,
\label{93c} 
\end{eqnarray}
$G^{ab}_{\mu\nu}(x,y,B) $ being the gluon propagator in the
background field (\ref{11}).

We can easily see that $W^{(2)}$ contains 3 pieces, two of them
are perturbative self-energy quark terms. The third one, assuming that
$x$ and $y$ refer to the  quark and antiquark trajectory
respectively, is the color Coulomb term, modified by the confining
background. As argued in Ref. \cite{17}, this term represents
the gluon propagating inside the world-sheet of the string between
$q$ and $\bar q$. When the time $T$ is large, the long film of
this world-sheet does not influence the motion of the gluon,
reducing it to the free OGE term. Hence $\lan W^{(2)}\ran$
factorizes into  the film term $(\lan W(B)\ran_B)$ and gluon
propagator $dx_\mu dy_\nu G_{\mu\nu}$, yielding finally
 the color Coulomb term in the potential. (This is however only true for 
the lowest order term $W^{(2)}$ and only  at large distances $|x-y|\la T_g$).
 Otherwise perturbative-nonperturbative interference comes into
 play\cite{34}.

 In what follows we restrict our attention to the first term, $\lan
 W(B)\ran_B$.
 Our next approximation is the neglect of perturbative exchanges in
 $\langle W\rangle $ 
 (they will be restored in the final expression for Hamiltonian).
 This yields for large Wilson loops, i.e. $R,T\gg T_g$,
 \be
 \langle W\rangle _B=const~ \exp(-\sigma S_{min})
 \label{94}
 \ee
 where $S_{min}$ is the minimal area inside the given trajectories 
$z(t), \bar z(t)$ of
 the quark and antiquark,
 \be
 S_{min} =\int^T_0 dt\int^1_0 d\beta \sqrt{\det g},~~
 g_{ab}=\partial_aw_\mu\partial_bw^{\mu},~a,~b=t,~\beta.
 \label{95}
 \ee
Here a point $w$ on the surface is parameterized by 
$w_\mu= \beta z_\mu(t)+(1-\beta) \bar z_\mu(t)$.

 The Nambu-Goto form of $S_{min}$ cannot be quantized due to the
 square root. To get rid of the square root  we may  use the auxiliary 
field approach\cite{42} with functions $\nu(\beta,t)$ and $\eta(\beta,t)$ 
as is usually done in string theories. As a result the
 total Euclidean action becomes\cite{33}
 \begin{eqnarray}
 A=&&K+\bar K+\sigma S_{min}=
 \int^T_0 dt\int^1_0 d\beta
 \{\frac12(\frac{m^2}{\mu(t)}+\frac{\bar m^2}{\bar
 \mu(t)})+\frac{\mu_+(t)}{2}\dot R^2
\nonumber
\\
&& +\frac{\tilde \mu(t)}{2}\dot r^2+
 \frac{\nu}{2}[\dot w^2+(\frac\sigma\nu)^2r^2-2\eta (\dot
 wr)+\eta^2r^2]\}.
 \label{96}
 \end{eqnarray}
 Here $\mu_+=\mu+\bar
\mu,~~\tilde\mu=\frac{\mu\bar\mu}{\mu+\bar\mu},~~
 R_i=\frac{\mu z_i+\bar\mu\bar z_i}{\mu+\bar\mu},~~
 r_i=z_i-\bar z_i$. 
 Performing the Gaussian integrations over $R_{\mu}$ and $\eta$ we
 arrive in the standard way at the Hamiltonian (we take $m=\bar m$
 for simplicity)
\begin{eqnarray} 
H =&&\frac{p^2_r+m^2}{\mu(\tau)}+ \mu(\tau)
\nonumber
\\
&&+\frac{\hat L^2/r^2}
{\mu+2\int^1_0(\beta-\frac{1}{2})^2\nu(\beta) d\beta}+
\frac{\sigma^2 r^2}{2}\int^1_0\frac{d\beta}{\nu(\beta)}+
\int^1_0\frac{\nu(\beta)}{2}d\beta, 
\label{17.a} 
\end{eqnarray}
where $p^2_r=(\vep\ver)^2/r^2$ and $L$ is the angular momentum, $\hat
L=(\ver\times \vep)$.

A reasonable approximation to the integrations over $\mu$ and $\nu$ is to
replace them by their corresponding extremum values\cite{33}.
For these values the terms $\mu(t)$ and $\nu(\beta)$ have a
simple physical meaning. E.g. when
$\sigma=0$ and $L=0$, we find from Eq. (\ref{17.a}) 
\be
H_0=2\sqrt{\vep^2+m^2},~~\mu_0=\sqrt{\vep^2+m^2},
\label{97} 
\ee 
so that $\mu_0$ corresponds to the energy of the quark. Similarly in the limiting
case $L\to \infty$ the extremum over $\nu(\beta)$ yields
\be
\nu_0(\beta)=\frac{\sigma r}{\sqrt{1-4y^2(\beta-\frac{1}{2})^2}},
~~H^2_0=2\pi\sigma\sqrt{L(L+1)}.
\label{98} 
\ee 
Hence $\nu_0$ is
the energy density along the string with $\beta$ playing the
role of the coordinate along the string.

\subsection{Nonperturbative fields as a correction}

Now we turn to the first type of expansion
 mentioned above, i.e. when the NP contribution is
 considered to be a (small) correction to a basically
 perturbative
 result. As an example let us consider the spectrum of heavy
 quarkonia. We can calculate the NP shift of the Coulombic levels of
 heavy quark--antiquark ($q\bar q$) system, following 
Ref. \cite{35}. When the quark mass $m$ is large, the spatial and
 temporal extensions of the n-th bound state are
 \be
 \bar r_n\simeq \frac{n}{m\alpha_s},~~\bar t_n\simeq
 \frac{n^2}{m\alpha^2_s}.
 \label{5.1}
 \ee
For low $n\sim 1$ these may be small enough to
 disregard the NP interaction in first approximation.
 So for the spin--averaged spectrum we can write
 \be
 M(n,l)=2m\{1-\frac{C_F\alpha_s^2}{8n^2}+0(\alpha^3_s)+\Delta_{NP}\}
 \label{5.2}
 \ee
 where
 $\Delta_{NP}$ is the expected nonperturbative correction, which
 should be small for states of small spatial extension.
This  conclusion can be drawn from the lattice (and phenomenological)
parameterization  of the static $q\bar q$ potential
 \be
 V(r)=-\frac{4\alpha_s(r)}{3r}+\sigma r + const
 \label{5.3}
 \ee

Using the empirical values found for $\sigma=0.2 GeV^2$ and 
$\alpha_s(r)\sim 0.3$ (at $r\approx 0.2 fm $)
we may  deduce that the first term on
the l.h.s. of Eq. (\ref{5.3}) is at $r\approx 0.3 fm$ 
comparable in magnitude to the second term, the NP contribution.
Hence this suggests that the states with a
radius $r\ll 0.3 fm$ are mainly governed by the (color) Coulomb
dynamics, while those with $r\gg 0.3 fm$ are mostly NP states. So we
may expect e.g. the $n=1$ bottomonium state to be largely Coulombic.

Let us now consider the general path--integral formalism for the $q\bar{q}$ 
system interacting via perturbative gluon exchanges and nonperturbative
correlators.
We start with the quark Green's function in the FSR form (cf. Eqs. (\ref{9},\ref{10}))
 \be
S(x,y)= i(m-\hat{D})\int^{\infty}_0 ds (Dz)_{xy} e^{-K}\Phi_\sigma(x,y),
 \label{5.4}
 \ee
 where 
 $$ K=m^2s+\frac{1}{4}\int^s_0\dot{z}^2_{\mu}d\tau $$ 
and $\Phi_\sigma$ contains spin insertions into the parallel transporter
 \be
 \Phi_\sigma (x,y) = P_AP_F \exp [ig \int^x_y A_{\mu} dz_{\mu}+ g \int^s_0 d\tau
 \sigma_{\mu\nu} F_{\mu\nu}(z(\tau))].
\label{5.5}
 \ee
Double ordering in $A_{\mu}$ and $F_{\mu\nu}$ is implied by the 
operators $P_A,P_F$.
We have  also introduced the $4\times 4$ matrix in Dirac space 
 \be
 \sigma_{\mu\nu} F_{\mu\nu} \equiv\vec{\sigma}_i
 \left (
 \begin{array}{cc}
 \vec{B}_i& \vec{E}_i\\
 \vec{E}_i&\vec{B}_i
 \end{array}
 \right )   .
 \label{5.6}
 \ee
Neglecting spins, we have instead of Eq. (\ref{5.5}) 
\be
\Phi_\sigma(x,y) \to \Phi(x,y) \equiv P_A 
\exp (ig \int^x_y A_{\mu}) dz_{\mu}. 
\label{5.7}
 \ee 
In terms of the single quark Green's functions (\ref{5.4}) and initial
and final state matrices $\Gamma_i, \Gamma_f$ ( such that
$\bar{q}(x)\Gamma_f\Phi(x,\bar{x}) q(\bar{x}) $ is the final
$q\bar{q}$ state) the total relativistic gauge--invariant
$q\bar{q}$ Green's function in the quenched approximation is
similar to Eq. (\ref{14})
\begin{eqnarray}
G(x,\bar{x}; y,\bar{y})&=&<tr (\Gamma_fS_1(x,y)\Gamma_i\Phi(y,\bar{y})
S_2(\bar{y},\bar{x})\Phi(\bar{x},x))|>
\nonumber
\\
&&-<tr(\Gamma_fS_1(x,\bar{x})\Phi(\bar{x},x))
tr(\Gamma_iS_2(\bar{y},y)\Phi(y,\bar{y}))>.
 \label{5.8}
 \end{eqnarray}
The angular brackets
in Eq. (\ref{5.8}) imply averaging over the gluonic field $A_{\mu}$
and the trace is taken over Dirac space.

Since we are interested in this case primarily in heavy quarkonia, it is
reasonable to do a systematic nonrelativistic approximation. To
this end we introduce as in Refs. \cite{36,33} the real evolution
(time) parameter $t$ instead of the proper time $\tau$ in
$K,(\bar{\tau}$ in $\bar{K}$) and the dynamical mass parameters
$\mu,\bar{\mu}$ as in Eq. (\ref{89})
 \be
  \frac{dt}{d\tau}= 2\mu_1,
\frac{dt}{d\bar{\tau}}=2\mu_2;~~
\int^s_0\dot{z}_{\mu}^2(\tau)d\tau=
\int^T_0 2\mu_1~dt(\frac{dz_{\mu}(t)}{dt})^2.
\label{5.9}
 \ee
Here we have denoted \be T\equiv \frac{1}{2}(x_4+\bar{x}_4).
\label{5.10}
 \ee
The nonrelativistic approximation is obtained, when we write for
$z_4(t),\bar{z}_4(t)$
\be 
z_4(t)=t+\zeta(t);~~~ \bar{z}_4(t)=t+\bar{\zeta}(t)
\label{5.11} 
\ee
and expands in the fluctuations $\zeta, \bar{\zeta}$, which are
$0(\frac{1}{\sqrt{m}})$. Note that the integration in $ds_1 ds_2$ goes
over into $d{\mu}_1d\mu_2$. Physically the expansion (\ref{5.11})
means that we neglect trajectories with backtracking of $z_4,
\bar{z}_4$, i.e. dropping the so-called $Z$ graphs.  We can persuade
ourself that the insertion of Eq. (\ref{5.11}) into $K,\bar{K}$ allows to
determine $\mu_1,\mu_2$ from the extremum in $K,\bar{K}$.
We get
\be
\mu_1=m_1+0(1/m_1), ~~\mu_2=m_2+0(1/m_2) .
\label{5.12}
 \ee
We can further make a systematic expansion in powers of
$1/m_i$\cite{36}).  At least to lowest orders in $1/m_i$ this
 procedure is equivalent to the standard (gauge--noninvariant)
 nonrelativistic expansion\cite{32}.

 Let us keep the leading term     of this expansion
 \be
 G(x\bar{x}, y\bar{y})= 4m_1m_2e^{-(m_1+m_2)T}\int D^3zD^3\bar{z}
 e^{-K_1-K_2} <W(C)> ,
 \label{5.13}
 \ee 
where $K_1=\frac{m_1}{2}\int^T_0\dot{z}_i^2(t)dt,~~
 K_2=\frac{m_2}{2}\int^T_0\dot{\bar{z}}_i^2(t)dt$. Furthermore,
$<W(C)>$ is the Wilson loop operator with a closed contour $C$ 
comprising the 
 $q$ and $\bar{q}$ paths, and the initial and final state parallel transporters
 $\Phi(x,\bar{x})$ and $\Phi(y,\bar{y})$.

 The representation (\ref{5.13})
 will be our main object of study in the remaining part of this Section.
 For the heavy $q\bar{q} $ system the perturbative interaction contains 
an expansion in
 powers of $\frac{\alpha_s}{v} (v$ being the velocity in the c.m. 
system). This  should  be
  kept entirely, while the nonperturbative interaction can be treated 
up to the lowest order approximation.
For the total gluonic field $A_{\mu}$ we may write
 \be
 A_{\mu}=B_{\mu}+a_{\mu},
 \label{5.14}
 \ee
 where $B_{\mu}$ is the NP background, while $a_{\mu}$ is the perturbative
 fluctuation.

 It is convenient  in this {\it first type of expansion} to split 
the gauge transformation as
 \be
 B_{\mu}\to V^+B_{\mu}V,~~
 a_{\mu}\to V^+ (a_{\mu}+\frac{i}{g}\partial_{\mu})V ,
 \label{5.15}
  \ee
 so that the parallel transporter
 \be
 \Phi(a;x,y)\equiv P exp (ig \int^x_y a_{\mu}dz_{\mu})
 \label{5.16}
 \ee
 transforms as
 \be
 \Phi(a;x,y)\to V^+(x)\Phi(a;x,y)V(y).
 \label{5.17}
 \ee

\begin{figure}[ht]
\begin{center}
\epsffile{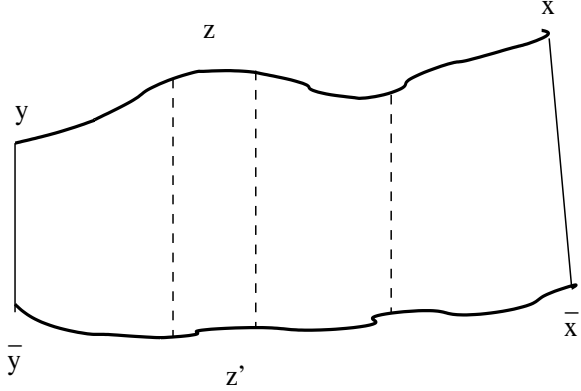}
\vspace{0.5cm}

\caption{\it The contour C, characterized by the quark trajectories $z$ and $z'$ 
in the Wilson loop with possible ladder-type  gluon exchanges between 
the quarks.}
\label{fig1}
\end{center} 
\end{figure}

Using Eq. (\ref{5.13}) we may now determine the effects of the NP
contribution as a correction.
The Wilson loop average in Eq. (\ref{5.13}) can be written, using 
Eq. (\ref{5.14}), as
\begin{eqnarray}
 <W&&(C)> \equiv <tr P exp (ig \int_C A_{\mu}dz_{\mu})>
 = <tr P exp (ig \int_C a_{\mu}dz_{\mu})>
\nonumber
\\
 &&+\frac{(ig)^2}{2!}\int_Cdz_{\mu}\int_Cdz'_{\mu}
  <tr P \Phi(a;z,z')B_{\nu}(z')\Phi(a;z',z) B_{\mu}(z)>
+~...
\nonumber
\\
&&= W_0+W_2+~...,
 \label{5.18}
\end{eqnarray}
 where we
 have omitted the term linear in $B_{\mu}$ since it vanishes when averaged
over the field $B_{\mu}$. The dots imply terms of higher power in
$B_{\mu}$.
The contour and points $z, z'$ are schematically shown in Fig. \ref{fig1}.

Let us discuss the first term on the r.h.s. of Eq. (\ref{5.18}). It is
the Wilson loop average of the usual perturbative fields, discussed
extensively in Ref. \cite{27}. We can use for $W_0$ the cluster
expansion to obtain  
\be W_0=Z \exp
(\varphi_2+\varphi_4+\varphi_6+...), \label{5.19}
\ee
\be 
\varphi_2\equiv -\frac{g^2}{8
\pi^2}\int\int\frac{dz_{\mu}dz'_{\mu}}{(z-z')^2}C_2,
~~C_2=\frac{N^2_c-1}{2N_c}
\label{5.20}
\ee
where regularization is implied in the integral $\varphi_2$ to be absorbed
into the $Z$ factor. Note that $\varphi_2$ contains all
ladder--type exchanges. 
In addition also
the "Abelian--crossed" diagrams --- those where times of the vertices can
not be ordered while color generators $t^a_i$ are always kept in the
same order, as in the ladder diagrams.
Therefore all crossed 
diagrams (minus "Abelian--crossed") are contained in $\varphi_4$
and contribute $0(1/N_c)$ as compared with ladder ones ($cf$
the discussion in Ref. \cite{27}). In addition $\varphi_4$ contains
"Mercedes--Benz diagrams", again repeated
infinitely many times. It is interesting to note that each term
$\varphi_2,\varphi_4$ etc. in Eq. (\ref{5.19}) sums up to an infinite
series of diagrams.

In particular $exp(\varphi_2)$ contains all terms with powers of
$\frac{\alpha_s}{v}$, as we shall see below.
For heavy (and slow) quarks we can write: 
\begin{eqnarray} 
\varphi_2&&=
\frac{g^2}{4\pi^2}\int^T_0\int^T_0
\frac{dtdt'C_2(1+\dot{\vec{z}}_i\dot{\vec{z}}'_i)}{\vec{r}^2+(t-t')^2}
\nonumber
\\
&&\approx
\int^T_0\frac{C_2\alpha_s}{|\vec{r}|}(1+0(v^2/c^2)dt=\varphi_2^{(0)}+0(v^2/c^2)
\label{5.21}
\end{eqnarray}
where $\vec{r}=\vec{z}-\vec{z}'$. In this way we obtain a
singlet one--gluon--exchange (OGE) potential, the effective time
difference being $\Delta t=|t-t'|\sim |\vec{r}|$.
In addition also the radiative corrections due to
the transverse gluon exchange can be obtained in this way.
For the $q\bar{q}$ mass $\varphi_2$ leads to a correction
of order $0(\alpha_s)$, being the Coulombic energy. It should however
be noted that in the wave--function 
the Coulomb potential has to be kept to all orders because of its
singular character. In fact we shall not expand
$exp (\varphi_2^{(0)})$, while this is done for
the other contributions like the  radiative corrections.

Turning to $W_2$ we may determine the leading (in $N_c$) set of
diagrams. They consist of the diagrams, where the gluons propagate
between the $q$ and $\bar{q}$ lines with the same time coordinates,
i.e. diagrams with Coulombic or instantaneous gluon exchanges. We have
for the gluon propagator
 \be
 <a_{\mu}^a(x)a_{\nu}^b(y)>=\frac{\delta_{ab}\delta_{\mu\nu}}{4\pi^2(x-y)^2}.
 \label{5.22}
   \ee
 Here we have chosen for simplicity the Feynman gauge, since $W_0$ and $W_2$
 are gauge invariant.
 Expanding $W_2$ in Eq. (\ref{5.18}) in powers of $a_{\mu}$ we have
in view of Eq. (\ref{5.22}) terms typically of the form
\begin{eqnarray}
tr(t^{b_k}t^{b_{k-1}}...t^{b_1}t^{b_1}t^{b_2}...&&t^{b_k}t^{a_1}...t^{a_n}
 t^{a}B^a_{\mu}t^{a_n}...t^{a_1}B^b_{\mu}t^{b}...)
\nonumber
\\
&& \to (C_2)^k tr
 (t^{a_1}...t^{a_n}t^{a}B^a_{\mu}t^{a_n}...t^{a_1}B^b_{\mu}t^{b}...)
 \label{5.23}
 \end{eqnarray}
 Now due to the equality
 \be
 t^ct^at^c=-\frac{1}{2N_c}t^a,
 \label{5.24}
 \ee
 We obtain for all exchanges in the time interval between times of
 $B_{\mu}(z)$ and $B_{\nu}(z')$,
 a factor $(-\frac{1}{2N_c})$ instead of the factor $C_2$ for all the other
 exchanges. 

\begin{figure}[ht]
\begin{center}
\epsffile{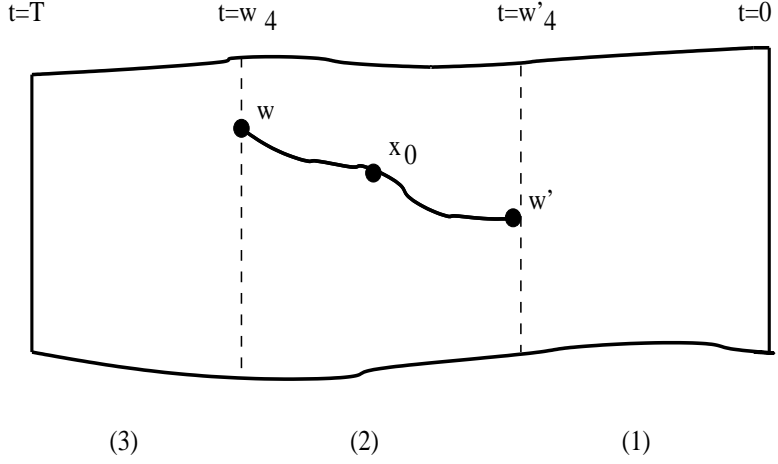}
\vspace{0.5cm}

\caption{\it 
Contribution of the Gaussian correlator to the $q\bar{q}$ Green's
function.The points w and w' of
the correlator are connected by the parallel transporter, shown by 
the solid line going through the point $x_0$. This makes the 
correlator gauge invariant.}
\label{fig2}
\end{center} 
\end{figure}

The correction term $W_2$ can be worked out explicitly.
In particular, we now derive the lowest order corrections to the energy 
levels and wave--functions due to the NP field correlators. 
As shown in Fig. \ref{fig2} 
we may divide the total time interval $T$ into three parts
  $$(1)~~~~0\leq  t\leq w_4'$$
   $$(2)~~~~w'_4\leq t\leq w_4$$
   $$(3)~~~~w_4\leq t\leq  T$$
where $t$ is the c.m. time. In $K+\bar{K}$ we may separate the c.m.
 and relative coordinates 
\be
 R_i=\frac{m_1z_i+m_2\bar{z}_i}{m_1+m_2};~~r_i(t)=z_i(t)-\bar{z}_i(t),
\label{5.31}
 \ee
 so that
 \be
 K+\bar{K}=\int^T_0\frac{M\dot{R}^2}{2}dt
 +\frac{\tilde{m}}{2}\int^T_0\dot{r}^2(t)dt,
\label{5.32}
 \ee
 with
 $\tilde{m}=\frac{m_1m_2}{m_1+m_2},~~M=m_1+m_2$.

  Separating out the trivial c.m. motion, we have in the parts (1) and (3)
  the path integrals, representing actually the singlet Coulomb Green's
  function
  \be
  G^{(1)}_C(r(t_1),r(t_2);t_1-t_2)=
  \int D\vec{r}(t)
     e^{-\frac{\tilde{m}}{2}\int^{t_1}_{t_2}
     \dot{\vec{r}}^2dt+C_2\alpha_s\int^{t_1}_{t_2}
     \frac{dt}{|r(t)|}}.
     \label{5.33}
          \ee
        In the part (2) instead we have an octet Coulomb Green's
     function
     \be
  G^{(8)}_C(r(w_4),r(w_4');w_4-w_4')=
  \int D\vec{r}(t)
     e^{-\frac{\tilde{m}}{2}\int^{w_4}_{w'_4}
     dt\dot{\vec{r}}^2-\frac{C_2\alpha_s}{2N_C}\int^{w_4}_{w'_4}
     \frac{dt}{|\vec{r}(t)|}} .
\label{5.34}  
\ee
As a result $W_2 $ can be written as
 \be
 W_2=\frac{(ig)^2}{2}\int_Cdz_{\mu}\int_C dz_{\nu}'<tr
 B_{\nu}B_{\mu}>e^{\int^T_{t_2}dtV^0_C+\int_0^{t_1}dtV^0_C+\int^{t_2}_{t_1}V_C^8dt} ,
 \label{5.25}
 \ee
 where
 \be
 V^0_C=C_2\frac{\alpha_s}{r},~~
 V^8_C=-\frac{1}{2N_c}\frac{\alpha_s}{r} .
 \label{5.26}
 \ee
We can easily identify $-V^0_C$ and $-V^8_C$ as a singlet and octet
 $q\bar{q}$ potential, considered in Ref. \cite{35}. For $<tr B_{\nu}B_{\mu}>$
 we can use the modified Fock--Schwinger gauge
to obtain:
  \be
 B_{\mu}(z) =\int^z_{x_0} dw_{\rho}\alpha(w)F_{\rho\mu}(w)
 \label{5.27} \ee and 
 \be \int dz_{\mu}dz_{\nu}'<B_{\mu}(z)B_{\nu}(z')>= \int d \sigma_{\mu\rho}
 d\sigma'_{\nu\lambda}<F_{\mu\rho}(w)F_{\nu\lambda}(w')>,
 \label{5.28}
 \ee
 where we have introduced a surface element
 $d\sigma_{\mu\rho} = dz_{\mu}dw_{\rho}\alpha(w)$.
 To make Eq. (\ref{5.28}) fully gauge--invariant we can introduce in the
 integral in Eq. (\ref{5.27}) factors $\Phi(x_0,w)$ identically equal to
 unity in the Fock--Schwinger gauge. As a result we get
 \begin{eqnarray}
  <tr B_{\nu}B_{\mu}>&&= <\int d\sigma_{ \mu\rho}d\sigma_{\nu\lambda}
\nonumber
\\
&&\times tr\{\Phi
 (x_0,w) F_{\mu\rho}(w)\Phi(w,x_0)
 \Phi(x_0,w')F_{\nu\lambda}(w')\Phi(w',x_0)\}> ,
 \label{5.30}
 \end{eqnarray}
which  is fully gauge invariant.

Using Eq. (\ref{5.25}) we find for the
    correction to the total $q\bar{q}$ Green's function 
$G=G^{(0)}+\Delta G,$ 
\begin{eqnarray} 
\Delta G=&&-\frac{g^2}{2}
\int d\sigma_{\mu\nu} (w) \int d\sigma_{\mu'\nu'}(w')
    d^3r(w_4) d^3r (w'_4) 
\nonumber
\\
&&G^{(1)}_c(r(T),r(w_4);T-w_4) 
G_c^{(8)}(r(w_4),r(w'_4),w_4-w'_4)
\nonumber
\\
&&~~~~~~~~\times <F_{\mu\nu}(w) F_{\mu'\nu'}(w')> 
G_c^{(1)}(r(w'_4),r(0);w'_4),
\label{5.35}
\end{eqnarray}
where the integrals over $d\sigma_{\mu\nu},~ d\sigma_{\mu'\nu'}$ are 
taken over the surface $\Sigma_{\mu_\nu}$.
It is convenient to identify $\Sigma_{\mu\nu} $ with the minimal surface
    inside the contour $C$, formed between the trajectories $z(t)$ and 
$\bar{z}(t)$. Introducing the straight line between $z(t)$ and 
$\bar{z}(t)$
    \be
    w_{\mu}(t,\beta) = z_{\mu}(t) \beta +\bar{z}_{\mu}(t) (1-\beta) =
    R_{\mu}+r_{\mu}(\beta -\frac{m_1}{m_1+m_2})
\label{5.36}
        \ee
with
    $$
    w_4 =z_4=\bar{z}_4=t
    $$
we may write the surface element as
    \be
    d\sigma_{\mu\nu}(w) =(w'_{\mu}\dot{w}_{\nu}-\dot{w}_{\mu}w'_{\nu}) 
dt
    d\beta \equiv a_{\mu\nu} dt{d\beta}
    \label{5.37}
    \ee
     with
      $w'_{\mu}=\frac{ \partial
    w_{\mu}}{\partial \beta}= r_{\mu},~ \dot{w}_{\mu}=\frac{\partial
    w_{\mu}}{\partial t}. $
     We also have in the c.m. system 
\be 
a_{i4}=r_{i};~~a_{ij}= e_{ijk} L_{k} 
\frac{1}{i\tilde{m}}(\beta-\frac{m_1}{m_1+m_2}),
    \label{5.38}
    \ee
while the Minkowskian angular momentum $L$ is given by
    \be
    L_i=e_{ikl}~r_k\cdot\frac{1}{i} \frac{\partial}{\partial r_l}.
    \label{5.39}
    \ee

    In the nonrelativistic approximation we expand in powers of
    $\frac{1}{m_1},\frac{1}{m_2}$. Hence $a_{ij}$ can be neglected in lowest
    order and we are left with only $a_{i4}$, i.e. in Eq. (\ref{5.35}) only
the electric field correlators should be kept. The field correlators 
have the following representation in terms of the two Lorentz 
invariants $D$ and $D_1$\cite{38}
   \begin{eqnarray}
    g^2tr&&<E_i(w)E_k(w')>=
\nonumber
\\
&&  \frac{1}{12}[\delta_{ik}(D(w-w')+D_1(w-w')+
  h^2_4\frac{\partial D_1}{\partial h^2})+
  h_ih_k\frac{\partial D_1}{\partial h^2}],
  \label{5.40}
\end{eqnarray}
  where $h_i=w_i-w'_i$.  $D$ and $D_1$ are normalized as
  \be
  D(0)+D_1(0)=g^2<tr F^2_{\mu\nu}(0)>=
  \frac{1}{2}4\pi^2 G_2.
  \label{5.41}
  \ee
$G_2$ is the standard definition of the gluonic condensate\cite{39}
  \be
  G_2 = \frac{\alpha_s}{\pi}<F^a_{\mu\nu}F^a_{\mu\nu}>=0.012
  GeV^4.
\label{5.42}
  \ee
  Inserting  Eq. (\ref{5.40}) into Eq. (\ref{5.35}) and neglecting  
the terms $h_ih_k\sim 0(\frac{1}{\tilde{m}^2})$ we get
  \begin{eqnarray}
   \Delta
  G=-\frac{1}{24} G_C^{(1)} (r(T),r)&& d^3r G_C^{(8)} (r,r') d^3 r'
\nonumber
\\
&&   r_i d\beta dw_4 r'_id\beta' dw'_4 \Delta(w-w') G_c^{(1)} (r',r(0)) ,
   \label{5.43}
   \end{eqnarray}
   where we have defined
   \be
   \Delta(w-w') = D(w-w') +D_1(w-w')+h^2_4 \frac{\partial D_1}{\partial h^2} .
   \label{5.44}
   \ee
   Using the spectral decomposition for $G_c$
   \be
   G^{(1,8)}_c(r,r',t)=<r|e^{-H_c^{(1,8)}t}|r'>=\sum_n \psi_n^{(1,8)}(r)
   e^{-E_n^{(1,8)}t}\psi_n^{(1,8)^+}(r'),
   \label{5.45}
   \ee
we can rewrite Eq. (\ref{5.35}) for the matrix element
    of $\Delta G$ between singlet Coulomb wave functions
    \begin{eqnarray}
    <n|\Delta&& G|n> = - \frac{e^{-E_n^{(1)}T}}{24}T \int \frac{dp_4
    d\vec{p}}{(2\pi)^4} \tilde{\Delta} (p) d\beta d\beta'
\nonumber
\\
&&\sum_{k=0,1,...}
 \frac{<n|r_ie^{i\vec{p}(\beta-\frac{m_1}{m_1+m_2})\vec{r}}|k>
<k|r'_i
e^{-i\bar{p}(\beta-\frac{m_1}{m_1+m_2})\vec{r}'}|n>}{E_k^{(8)}-E_n-ip_4},
    \label{5.46}
    \end{eqnarray}

where $\tilde{\Delta}(p)$ is the Fourier transform of Eq. (\ref{5.44}).
The set of states $|k>$ in Eq. (\ref{5.46}) with eigenvalues $E_k^{(8)}$ refer to the
    octet Hamiltonian piece in Eq.  (\ref{5.34})
    \be
     H^{(8)} =
    \frac{\vec{p}^2}{2\tilde{m}}+\frac{C_2\alpha_s}{2N_c|\vec{r}|} .
    \label{5.47}
    \ee

    The correlator $\Delta(x)$ depends  on $x$ as
    $\Delta (x)= f (\frac{|x|}{T_g})$, and decays exponentially at large
    $|x|$\cite{33}. For what follows it is crucial to compare the two
    parameters,  $T_g$ and  the Coulombic size of the $n$-th state of
    the  $q\bar{q}$ system,  $R_n=\frac{n}{\tilde{m}c_2\alpha_s}$.
    In the Voloshin-Leutwyler case\cite{36} it is assumed explicitly 
or implicitly that $${\rm Case~(i)}~~ T_g\gg R_n$$
    In the opposite case
    $${\rm Case~(ii)}~~ T_g\ll R_n$$
    as we shall see completely different dynamics  occurs.

    Writing $G_0+\Delta G=const~e^{-(E_n^{(1)}+\Delta E_n)T}\approx const~
    e^{-E_n^{(1)}T}(1-\Delta E_nT)$  we finally obtain in case (i)
for $\Delta E_n$
    \be
    \Delta E_n=\frac{\pi^2
    G_2}{18}\frac{<n|r_i|k><k|r_i|n>}{E_k^{(8)}-E_n}.
    \label{5.50}
    \ee
Results of the calculations~\cite{35} using the Voloshin-Leutwyler approximation
(VLA) (Eq. (\ref{5.50})
for charmonium and bottomonium) and using the 
standard value of $G_2$\cite{39} are given in Table 1. One can see that a rough
agreement exists only for the lowest bottomonium state. Consider
now the opposite case, $T_g\ll R_n$. Since lattice measurements
yield $T_g\approx 0.2 fm$, we may expect that this case is
generally applicable to all $b\bar b$ and $c\bar c$ states.
However in this case it is not enough to keep only the $w_2$, but
also sum up all NP terms, which amounts to the exponentiation of
the NP contribution. The NP local potential appears in
addition to the Coulomb term.
 This has been done in the framework of the local  potential picture in
 Ref. \cite{41}, where the NP potential is expressed via  correlators  
$D(x)$ and $D_1(x)$.

Results of calculations of Ref. \cite{41} yield a very
 consistent picture both for levels and wave functions of bottomonium
 and quarkonium. To compare with VLA  and our results here the
 results of Ref. \cite{41} are listed in the middle column of the Table,
 demonstrating a much better agreement with experiment than results for
the  VLA.
 Note, that in  Ref. \cite{41} the NP interaction  was not treated  as a
 perturbation, but nonperturbatively by including
the NP part in the potential. Consequently this
explains why the predictions have improved.
 To summarize, because  of the small  $T_g\approx 0.2 fm$, the
 potential picture is more adequate for quarkonia than the VLA
 formalism or QCD sum rules, including even the bottomonium
 case.

\begin{center}
{\bf Table 1.\\}
\vspace{0.1in}

The experimental values and the predicted splittings in $MeV$ of  various 
states in bottomium and charmonium in the Voloshin-Leutwyler approximation 
and Ref. \cite{41}.
\vspace{0.2in}

\begin{tabular}{llll}
\hline
splitting& VLA &\cite{41}  &exp.\\
\hline
$2S-1S(b\bar b)$  &479 &554  &558 \\
$2S-2P(b\bar b)$  &181 &112  &123 \\
$3S-2S(b\bar b)$  &4570 &342 &332 \\
$2S-1S(c\bar c)$  &9733 &582  &670 \\
\hline
\end{tabular}
\end{center}
\vspace{1.cm}

\section{ IR and collinear singularities in FSR}

It is known that in QED some matrix elements and partial
cross-sections display singularities\cite{43}, which are of two
general types; a) due to  soft photon exchange (IR singularities)
b) due to the collinear motion of an exchanged and  emitted photon
(collinear singularities).
A similar situation exists in perturbative QCD, where the same
lowest order amplitudes contain both IR and collinear
singularities, see Ref. \cite{44}.

\begin{figure}[ht]
\begin{center}
\epsffile{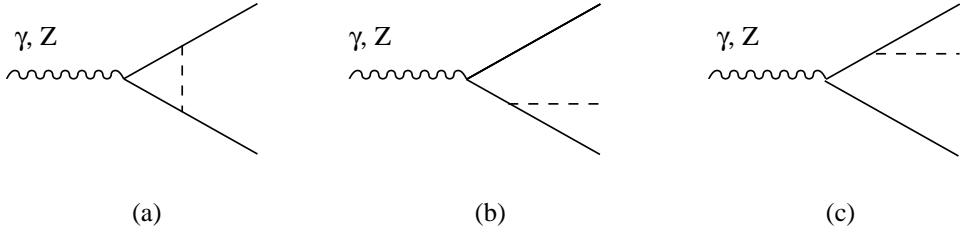}
\vspace{0.5cm}

\caption{\it Feynman graphs to order $\alpha_s$ with singularities,
 which are cancelled in the crossection.}
\label{fig3}
\end{center}
\end{figure}
Examples of Feynman graphs which produce both types of
singularities 
are given in Fig. \ref{fig3}. These graphs 
refer to the process $e^+ e^- \to \bar q q$ and the
total cross section for the sum of three graphs where an
additional gluon can be generated is finite to the order
$O(\alpha_s)$ and is equal to 
\be 
\sigma^{(1)} = \sigma^{(0)}
\left \{ 1+\frac{3C_F\alpha_s}{4\pi}\right\}. 
\label{8.1} 
\ee 
This is in line with the KLN theorem\cite{45} derived and proved in
the framework of QED.

\begin{figure}[ht]
\begin{center}
\epsffile{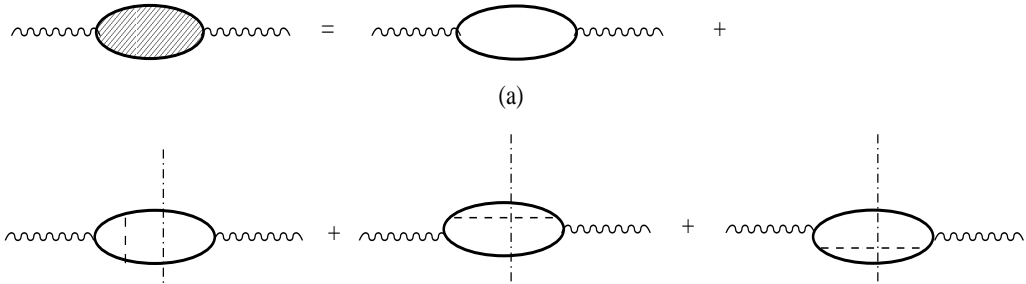}
\vspace{0.5cm}

\caption{\it The gauge-invariant $\gamma-\gamma$ amplitude,
where all
sigularities cancel, while they are separately present in the parts of graphs
obtained by cutting along the dash-dotted lines.}
\label{fig4}
\end{center}
\end{figure}

 We now argue that the situation is  different in QCD
when the nonperturbative confining vacuum is taken into account. 
This in particular can be demonstrated in the framework of the FSR.
We start with the QCD contribution to the photon self-energy part,
which is shown graphically in Fig. \ref{fig4}. It has the form 
 \be 
\Pi_{\mu\nu} (Q) =
(Q_\mu Q_\nu- Q^2 \delta_{\mu\nu}) \Pi (Q^2) .
\label{8.2} 
\ee 
In particular, the $O(\alpha_s)$ part  of $\Pi_{\mu\nu}(Q^2)$,
shown in Fig. \ref{fig4}b,
yields an analytic function of $Q^2$, with  the
imaginary part  (discontinuity across the cut $Q^2\geq 0$) given
by the sum of the three graphs
of Fig. \ref{fig4}a,b multiplied by the complementary parts as shown in
by the dash-dotted lines in Fig. \ref{fig4}b.
It is clear, that both the whole function $\Pi(Q^2)$ and its
total absorptive part is free from IR and collinear singularities,
while each piece in  the imaginary part, yielding partial
crossections $\sigma_{(a)},~ \sigma_{(b,c)}$
corresponding to the graphs Fig. \ref{fig3}a-c 
are IR divergent.  At this point the
difference between QED and QCD can be felt even on the purely
perturbative level. Namely, in QED the process $e^+e^-\to
e^+e^-+n\gamma$ cannot be associated with the imaginary part of
photon self-energy $\Pi_\gamma (Q^2)$, since photons can  be
emitted in  any amount off the $\Pi_\gamma (Q^2)$ and hence should
be summed up separately.

 This fact is formulated as a notion of  a physical electron,
 containing a bare electron plus any amount of additional soft photons 
(see the Bloch-Nordsiek method discussed for example in Ref. \cite{43}).
 In QCD the situation is different, since separate gluons cannot
 escape the internal space of $\Pi_{\mu\nu} (Q^2)$ (cannot be
 emitted), except when they create (pairwise, triplewise etc.)
 massive glueballs, or else when they are accompained by the sea
 quark pairs forming hybrid states.

 To illustrate
 our ideas we shall use the FSR, introducing the background confining
 field,  and using the method of Ref. \cite{46}.
We define the photon vacuum polarization function $\Pi(q^2)$ as a
correlator of electromagnetic currents for the process $e^+e^-\to
$ hadrons in the usual way 
\be 
-i\int d^4xe^{iqx}<0|T(j_{\mu}(x)j_{\nu}(0)|0>= (q_{\mu }
q_{\nu}-g_{\mu\nu}q^2)\Pi(q^2),
\label{20f}
 \ee 
where the imaginary
part of $\Pi$ is related to the total hadronic ratio $R$ as 
\be
R(q^2)=\frac{\sigma(e^+e^-\to
hadrons)}{\sigma(e^+e^-\to\mu^+\mu^-)}= 12\pi Im
\Pi(\frac{q^2}{\mu^2},\alpha_s(\mu)). \label{21f} 
\ee

 There are two usual approaches to calculate $\Pi(q^2)$.
The first one is based on a purely perturbative expansion,
which is now known to the order $0(\alpha^3_s)$\cite{47}. The
second one is the OPE approach\cite{39}, which includes the
NP contributions in the form of local condensates. For
two light quarks of equal masses $(m_u=m_d=m)$ it
yields for $\Pi(Q^2)$
 \be
\!\!\!\!\!\!\!\Pi(Q^2)=-\frac{1}{4\pi^2}(1+\frac{\alpha_s}{\pi})ln
\frac{Q^2}{\mu^2}+\frac{6m^2}{Q^2}+\frac{2m<q \bar q>}{Q^4}
+\frac{\alpha_s<FF>}{12\pi Q^4}+... 
\label{22f}
 \ee 
In what follows we shall include the NP fields as they enter into Green's
functions, i.e. nonlocally. Moreover, we  shall be  mostly interested in
the large distance behaviour, where the role of NP fields is important.
To this end we first of all write the exact expression for
$\Pi(Q^2)$ in the presence of the nonperturbative background and
formulate some general properties of the perturbative series. 
More specifically, the e.m. current correlator can be written in 
the form\cite{17,46} 
\begin{eqnarray}
\Pi(Q^2)=&&\frac{1}{N}\int e^{iQx}d^4x\int DB \int
Dae^{-S_E(B+a)}
\nonumber
\\
&&\times tr(\gamma_{\mu}G_{q}(x,0) \gamma_{\mu}G_q(0,x))
det (m+\hat D(B+a))
\label{23f} 
\end{eqnarray}
Here $G_q$ is the single quark Green's function in the total field
$B_{\mu}+a_{\mu}$, \be G_q^{(B+a)}(x,y)=<x|(m+\hat{\partial}-ig
(\hat B+\hat a))^{-1}|y>.
\label{24f} 
\ee 
The background quark
propagator is conveniently written using the FSR as
 \be 
G^{(B)}_q(x,y)= \int^{\infty}_0 ds
e^{-K}DzP \exp[ ig\int^x_y (B_{\mu}) dz_{\mu}] \exp
[g\int^s_0\sigma_{\mu\nu}F_{\mu\nu} d\tau ]
\label{25f} 
\ee 
with $K=\frac{1}{4}\int^s_0\dot z^2(\tau) d\tau$.

\begin{figure}[ht]
\vspace{0.5cm}
\begin{center}
\epsffile{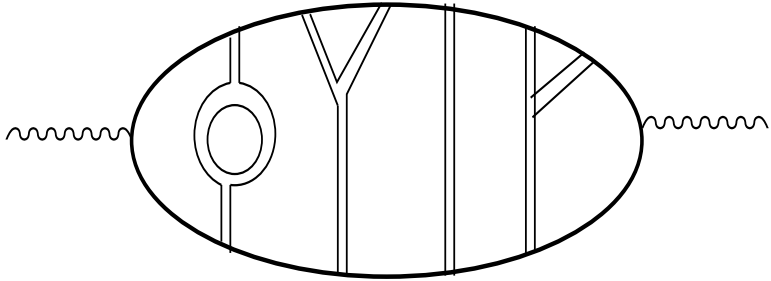}
\vspace{0.5cm}

\end{center}
\caption{\it A generic Feynman diagram for the photon self-energy in high
order of background perturbation theory
at large $N_c$. All areas $\Delta S_k$ between double gluon lines are
covered by the confining film, yielding the area law in Eq. (\ref{26f}).}
\label{fig5}
\end{figure}

 To simplify our analysis we take the limit
 $N_c\to\infty$  end drop the det term in Eq. (\ref{23f}). We are then left
 with only planar diagrams containing gluon exchanges
 $G_q^{(B)}$ in the external background field. Moreover,
 all gluon lines $G_q^{(B)}$ in the limit $N_c\to
 \infty$ are replaced by double fundamental lines\cite{48} and
 we are left only with diagrams, where the area $S$
 between the quark lines in $\Pi(Q^2)$ is divided into a
 number of pieces $\Delta S_k$,
shown schematically in Fig. \ref{fig5}. 

 The infrared behaviour of $\alpha_s(Q^2)$ at small
 $Q^2$ is connected to the limit of large areas of $S$.
In this limit the product of all phase integrals $\Phi(x_i,y_i)=
 exp (ig \int^{y_i}_{x_i} B_{\mu}dz_{\mu})$ from all
 Green's functions can be averaged using the area law,
 i.e. we have (modulo spin insertions $\sigma F$, which
 are unimportant for large distances\cite{17})
 \begin{eqnarray}
 <\Pi^n_{i=1}\Phi(x_i,y_i)>_B&&=
\Pi^n_{k=1}<W(\Delta S_k)>
\nonumber
\\
&&\approx\exp
 (-\sigma\sum^{ n}_{k=1} \Delta S_k) ~for~{N_c\to \infty}.
\label{26f}
 \end{eqnarray}
 This last factor serves as the IR regularizing  factor
 in the Feynman integral, preventing any type  of IR divergence.
 Using representations (\ref{23f}) and (\ref{25f}), we can formulate the following
 theorem.

\noindent
  \textbf{Theorem:}

   Any term in the perturbative expansion of $\Pi(Q^2)$, Eq. (\ref{23f}), in
   powers of $g a_{\mu}$ at large $N_c$ can be written as a
   configuration space Feynman diagram with an additional weight
   $<W_{s}(C_i)>$ for each closed contour.
   Here $W_{s}(C_i)$ is the Wilson loop with spin insertions,
   as in Eq. (\ref{25f}). Brackets  denote averaging over background fields.

   This theorem is easily proved by expanding  Eq. (\ref{23f}) in powers of $g
   a_{\mu}$ and using  double quark lines, Eq. (\ref{25f}), for gluon lines. Since at large $N_c$ we
   can replace adjoint color indices by doubled fundamental ones,
   then in each planar diagram the whole surface is divided into a
   set of closed fundamental contours, for which Eq. (\ref{26f}) holds true,
   again due to large $N_c$. Thus to each contour is assigned the
   Wilson loop $<W(C_i)>$. The rest is the usual free propagators
   written in the FSR.~

   Looking now at the large distance
   behaviour of the resulting configuration
   space planar Feynman diagram, we may derive from the
   above theorem
   the following corollary.

\noindent
   \textbf{Corollary:}

   Any planar diagram for $\Pi(Q^2)$ is convergent at large distances
   in the Euclidean space--time in the confining phase, when
   $<W(C_i)>\sim exp (-\sigma S_i)$.

   The proof is trivial, since the free planar diagram may diverge at
   large distances at most logarithmically.
   The kernel $<W(C_i)>$ makes all integrals convergent at large distances.
   At small distances (i.e. for small area, $S_i\to 0$) the kernel
   $<W(C_i)>$ behaves as  
$$<W(C_i)>\sim\exp (-\frac{g^2<F^a_{\mu\nu}F^a_{\mu\nu}(0)>S_i^2}{24N_c})$$ 
(see second reference in Ref. \cite{38}).
   Hence the
   structure of small-distance perturbative divergencies is the same
   for the planar diagram whether the NP background is present or
   not. Therefore the usual   renormalization technique (e.g. the
   dimensional renormalization) is applicable. As a result the
   planar Feynman diagram contributions to $\alpha_s^n\Pi^{(n)}(Q^2)$
   in the background is made finite also at small distances. The
   consequence of this is that any renormalized term
   $\alpha^n_s\Pi^{(n)}(Q^2)$ in the perturbative expansion of
   $\Pi(Q^2)$ is finite at all finite Euclidean $Q^2$, including
   $Q^2=0$.

   Then it follows that we can choose the renormalization scheme for
   $\alpha_s$, which renders $\alpha_s$ finite for all $0\leq
   Q^2<\infty$ and the Landau ghost pole will be absent.
   To make explicit this renormalization of $\alpha_s$, we can write
   the perturbative expansion of the function $\Pi(Q^2)$, Eq. (\ref{23f}), as
   \be
   \Pi(Q^2)=\Pi^{(0)}(Q^2)+\alpha_s\Pi^{(1)}(Q^2)+\alpha_s^2\Pi^{(2)}
   (Q^2)+...\label{27f}
   \ee
   We now again use the large $N_c$ approximation, in which case
   $\Pi^{(0)}$ contains only simple poles in $Q^2$\cite{48}:
   \be
   \Pi^{(0)}(Q^2)=\frac{1}{12 \pi^2}
   \sum^{\infty}_{n=0}\frac{C_n}{Q^2+M^2_n},
\label{28f}
   \ee
   where the mass $M_n$ is an eigenvalue of the Hamiltonian $H^{(0)}$.
It contains only quarks and background field $B_{\mu}$,
   \be
   H^{(0)}\Psi_n=M_n\Psi_n,\label{29f}
   \ee
   while the constant $C_n$ is connected to the eigenfunctions of
   $H^{(0)}$\cite{46}. We have
   \be
   C_n= \frac{N_cQ^2_ff_n^2\lambda_n^2}{M_n},\label{30f}
     \ee
   where
   $$
   f_n=\frac{1}{2\pi^2}\int^{\infty}_0u_n(k)kdk
   \frac{E_k+m}{E_k}(1+\frac{1}{3}
   \frac{(E_k-m)}{(E_k+m)}),$$
   $$
   \lambda^2_n=2\pi^2(\int^{\infty}_0dku^2_n(k))^{-1};~~
   E_k\equiv (k^2+m^2)^{1/2}.
   $$

      In what follows we are mostly interested in the long--distance
   effective Hamiltonian. It can be obtained from $G_{q\bar q}$ for large
   distances, $r\gg T_g$, where $T_g$ is the gluonic correlation
   length of the vacuum, $T_g\approx 0.2 fm$\cite{40}:
   \be
   G_{q\bar q}(x,0)=\int
   DB\eta(B)tr(\gamma_{\mu}G_q(x,0)\gamma_{\mu}G_q(0,x))=
   <x|e^{-H^{(0)}|x|}|0>.\label{31f}
   \ee
    At these distances we can neglect in Eq. (\ref{25f}) the  quark spin
    insertions $\sigma_{\mu\nu}F_{\mu\nu}$ and use the area law:
    \be
    <W_C>\to \exp (-\sigma S_{min}),
\label{32f}
    \ee
    where $S_{min}$ is the minimal area inside the loop $C$.

    Then the Hamiltonian in Eq. (\ref{31f}) is readily obtained by the method of
    Ref. \cite{42}. In the c.m. system for the orbital momentum $l=0$ it has the
    familiar form:
    \be
    H^{(0)}=2\sqrt{\vec p^2+m^2}+\sigma r + {\rm const},
\label{33f}
    \ee
    where a constant appears due to the perimeter term in $<W_C>$.
For $l=2$ a small correction from the rotating string
appears\cite{33}, which we neglect in first approximation.

    Now we can use the results of the quasiclassical analysis of
    $H^{(0)}$\cite{49}, where the values of $M_n,C_n$ have already been
    found.
    They can be represented as follows
    $(n=n_r+l/2,n_r=0,1,2,...,l=0,2)$
    \be
    M^2_n=2\pi\sigma (2n_r+l)+M_0^2,
\label{34f}
    \ee
    where $M_0^2$ is a weak function of the  quantum numbers $n_r,l$
    separately,
    comprising  the constant term of Eq. (\ref{33f}). In what follows we shall
    put it equal to the $\rho$--meson mass, $M_0^2\simeq m^2_{\rho}$.
     For $C_n$ one obtains quasiclassically\cite{49}
     $$
     C_n(l=0)=\frac{2}{3}Q^2_fN_cm^2_0,~~m_0^2\equiv 4\pi\sigma,
     $$
     \be
     C_n(l=2)=\frac{1}{3}Q^2_fN_cm^2_0.\label{35f}
     \ee
     Using the asymptotic expressions Eqs. (\ref{34f}-\ref{35f}) for $M_n,C_n$ 
and starting with $n=n_0$, we can write
     \begin{eqnarray}
     \Pi^{(0)}(Q^2)=\frac{1}{12\pi^2}\sum^{n_0-1}_{n=0}
     \frac{C_n}{M^2_n+Q^2} - &&
     \frac{Q^2_fN_c}{12\pi^2}\psi(\frac{Q^2+M_0^2+n_0m_0^2}{m_0^2})\label{36f}
\nonumber
\\
&& + ~divergent~~ constant.
     \end{eqnarray}
Here we have used the equality
\be
     \sum^{\infty}_{n=n_0}\frac{1}{M^2_n+Q^2}=
     -\frac{1}{m_0^2}
     \psi(\frac{Q^2+M_0^2+n_0m_0^2}{m_0^2})\label{37f}
     + ~divergent~~ constant
\ee
and $\psi(z)=\frac{\Gamma'(z)}{\Gamma(z)}$.

     In Eq. (\ref{36f}) we have separated the first $n_0$ terms to
     treat them nonquasiclassically, while keeping for the
     other states with $n\geq n_0$ the quasiclassical
     expressions (\ref{34f}-\ref{35f}). In what follows, however, we
     shall put $n_0=1$ for simplicity. We shall show
     below that even in this case our results will
     reproduce $e^+e^-$ experimental data with good
     accuracy (see Ref. \cite{46} for details).

Consider the asymptotics of $\Pi^{(0)}(Q^2)$ at large $Q^2$.
     Using the asymptotics of $\psi(z)$:
\be 
\psi(z)_{z\to\infty}=ln
~z-\frac{1}{2z}-\sum^{\infty}_{k=1}\frac{B_{2k}}{2kz^{2k}},
\label{38f}
\ee
where  $B_n$ are Bernoulli numbers, we obtain from Eq. (\ref{36f}) 
\be
\Pi^{(0)}(Q^2)=-\frac{Q_f^2 N_c}{12\pi^2}\ln
\frac{Q^2+M^2_0}{\mu^2}+0(\frac{m_0^2}{Q^2}).\label{39f}
\ee

We can easily see that this term coincides at $Q^2\gg M_0^2$ with
the first term in the OPE (\ref{22f}) -- the logarithmic one. Taking
the imaginary part of Eq. (\ref{39f}) at $Q^2\to -s$ we find 
\be
R(q^2)=12\pi Im\Pi^{(0)}(-s)=N_cQ^2_f,
\label{40f} 
\ee 
i.e. it
means that we have obtained for $\Pi^{(0)}$ the same result as for 
free quarks. This fact is the explicit manifestation of the
quark--hadron duality.

The  analysis of more complicated planar graphs as in Fig. \ref{fig5} 
can be carried out as in Ref. \cite{46}, yielding in this way a new
perturbative series with $\alpha_s$ renormalized in the background
fields and having therefore no Landau ghost poles. We 
refer to Refs. \cite{17} and \cite{46} for more details.

Now instead we return to the diagrams in Fig. \ref{fig3}, 
which give the lowest order perturbative amplitudes associated
with the production of 2 and 3 jets. 
The corresponding photon self-energy part is depicted in Fig. \ref{fig4}b.  
When the nonperturbative interaction is disregarded, even in the
hadronization process, so that a gluon emitted with the moment $k$
is directly associated with a gluon jet, then the singularities of
the partial cross sections
of Fig. \ref{fig3}
are transmitted into
the singularities of the jet cross sections. These are cured by
the introduction of the jet thickness, as in Sterman-Weinberg
method\cite{50} or introducing finite angular resolution
$\eta_0$ to distinguish 2-jet and 3-jet events (see Refs. 
\cite{44},\cite{51} for
details). In what follows we show, that in the FSR with account of
background fields, all IR and collonear singularities disappear.
Therefore the cross sections for 2-jet and 3-jet events remain
finite.

As was discussed above, in the leading $1/N_c$ approximation we
have only planar graphs for $\prod (Q^2)$, describing 
"1-jet events", which actually create a constant behaviour of
$R(q^2)$ (apart from new opening thresholds), exactly reproducing
the hadronic ratio, see Eq. (\ref{40f}).
Speaking of 2-jet events
 we may actually consider the next approximation in $1/N_c$, since we
 need an extra quark loop in the $\Pi(Q^2)$.
 This can be easily derived from Eq. (\ref{23f}), where the determinant
 can be expanded in the FSR
 \be
 \ln\det(m+\hat{D})=\frac{1}{2}\ln[\det(m^2-\hat{D}^2)]=
\frac{1}{2}tr\ln(m^2-\hat{D}^2),
\label{8.3} 
\ee
where we have used the symmetry property of the spectrum of $\hat{D}$.
 Hence 
\be
\det(m+\hat{D})=\exp\left\{-\frac12 tr\int\frac{ds}{s}\xi(s)
e^{-sm^2-K} Dz_{xx} W_\sigma(A,F) \right\}
\label{8.4} 
\ee 
with $\xi(s)$ a regularizing factor. For this we may take 
$\xi(t)= \lim \frac{d}{ds}\frac{M^{2s}t^s}{\Gamma (s)}|_{s=0}$ 
or we can use the Pauli-Villars form for $\xi(t)$.
Furthermore in Eq. (\ref{8.4}) we have 
\be
W_\sigma(A,F) = P_AP_F
\exp ig \int_C A_\mu dz_\mu\cdot \exp
g\int^s_0\sigma_{\mu\nu} F_{\mu\nu} d\tau. 
\label{8.5} 
\ee 
It is clear that $\det (m+\hat D)$ allows an expansion in the number of
quark loops, which is done by expanding the exponential in Eq. (\ref{8.4}).
Keeping only one quark loop for the 2-jet events, we obtain the
graph, shown in Fig. 6
%\ref{fig6} 
with an internal quark loop from the
determinant. It is essential, that the whole region between the
loops is covered by the NP correlators, creating a kind of "film"-
the world surface of the string, with perturbative (i.e. generated
by $a_\mu$) exchanges.

\begin{figure}[ht]
\begin{center}
\epsffile{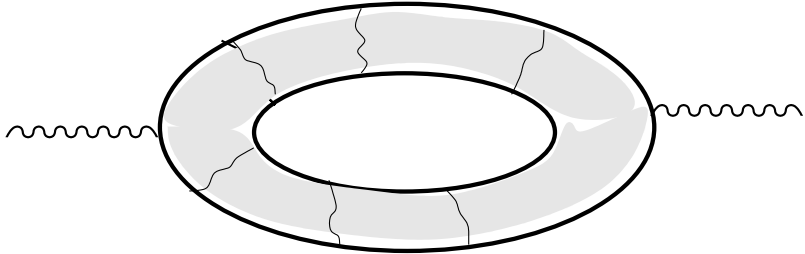}
\vspace{0.5cm}

\caption{\it Photon self-energy graph corresponding to the 2-jet cross-section
with one dynamical quark loop.}
\end{center}
\label{fig6}
\end{figure}

\begin{figure}[ht]
\begin{center}
\epsffile{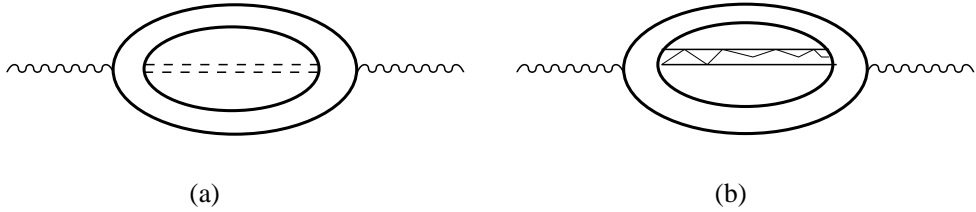}
\vspace{0.5cm}
\caption{\it Graphs corresponding to the 3-jet cross-section with the gluon
hadronized into a glueball or
accompanied by a sea-quark pair forming a hybrid.}
\label{fig7}
\end{center}
\end{figure}

It is clear that in this situation quarks are {\it never on the mass shell}
 (in contrast to the purely perturbative case) and therefore both
 IR and collinear singularities are absent.
 We can consider also the amplitude for  the 3-jet event, with one gluon jet.
Its perturbative amplitude corresponds to Fig. \ref{fig3}b,c. 
The perturbative situation is discussed in detail in Ref. \cite{51},
 and the 3-jet cross section for the process $e^+e^-$ 
$\to q\bar q g$ is given by 
\be
\frac{1}{\sigma} \frac{d^2\sigma}{dx_1dx_2} = 
C_F\frac{\alpha_s}{2\pi} \frac{x^2_1+x^2_2}{(1-x_1)(1-x_2)},
\label{205}
\ee
where the integration region is $0\leq x_1, x_2\leq 1,~~ x_1+x_2\geq 1.$
The integral is divergent both due to collinear and IR effects, since 
$1-x_1= x_2E_g(1-\cos \theta_{2g})/ \sqrt{s}$ and 
$1-x_2= x_1E_g(1-\cos \theta_{1g})/ \sqrt{s}$, where   $E_g$ is gluon energy 
and $\theta_{ig}$ is angle between the gluon and i-th quark.
 To handle these divergencies one can use the so-called JADE algorithm\cite{52},  
where the minimum invariant mass of a parton pair is larger than 
$ys$, i.e. $min (p_i+p_j)^2> ys$. With this condition the energy 
region for 3-jet events looks like
\be
0<x_1, x_2<1-y,~~ x_1+x_2>1+y.
\label{206}
\ee
The nonperturbative counterpart is obtained in two ways: a)
the emitted gluon is accompanied by another gluon, forming together
a two-gluon glueball (as was calculated in the framework of FSR
in Ref. \cite{53}), or b) a hybrid formation, when the emitted gluon is
accompanied by a sea quark-antiquark pair. 
Both possibilities are depicted in Fig. \ref{fig7}. 
For the hybrid case we should expand
the determinant term (the exponential in Eq. (\ref{8.4})) to the
second power, producing in this way the two quark loops.

It is clear in this case, that all particles, including
the gluon, are off-shell and IR and collinear singularities are
absent.  Moreover, assuming as usual almost collinear 
hadronisation\cite{51} we should replace the momenta of quarks 
and gluons by the corresponding momenta of hadrons. We can easily 
see that the factor
$\frac{1}{2p_1 k}$ is singular in case of $q\bar q g$ system becomes
$\frac{1}{2p_1k+\Delta M^2}$, where  $\Delta M^2 =M^2_1-M^2_2+M^2_g$ 
and $M_g$ is the hybrid (glueball) mass, so that $\Delta  M^2$ is
 in the GeV region. It effectively cuts off the singularity at small $y_{cut}$,
as is clearly seen in the experimental data\cite{54}.
 The same reasoning applies for higher jet events. The
only issue which exists is of experimental character. It amounts
to the precise definition of the number of jets, i.e. to classify the hadrons
between the several jets.

Thus  experiment as well as background perturbation 
theory do not exhibit collinear and IR singularities pertinent to the 
standard perturbation theory.

\section{FSR at nonzero temperature}

Within the framework of the FSR the problem of the various Green's functions
at finite temperature can be studied\cite{11,12}.
We first discuss the basic formalism for $T>0$ and then 
turn to the calculation of the gluon and quark  Green's functions.

\subsection{Basic equations}

We start with standard formulae of the background field 
formalism\cite{16,17} generalized to the case of nonzero temperature. 
We assume that the gluonic field $A_\mu$ can be split into the
background field $B_\mu$ and the quantum field $a_\mu$ 
\be 
A_\mu=B_\mu+a_\mu, 
\label{23a} 
\ee 
both satisfying the periodic  boundary conditions
\be 
B_\mu(z_4,z_i) =B_\mu(z_4+n\beta, z_i);~~a_\mu(z_4,z_i) =a_\mu(z_4+n\beta, z_i), 
\label{24a} 
\ee 
where
$n$ is an integer and $\beta=1/T$. The partition function can be
written as
\be 
Z(V,T) =\lan Z(B)\ran_B
\label{25a}
\ee
with
\be
Z(B)=N\int D\phi \exp \left(-\int^\beta_0 d\tau \int d^3x
L_{tot} (x,\tau)\right ) 
 \ee 
and where $\phi$ denotes all
set of fields $a_\mu,\Psi, \Psi^+$, $L_{tot}$ is the same as $L(a)$ defined 
in Eq. (\ref{la}) and $N$ is a normalization constant. 
Furthermore, in Eq. (\ref{25a}) $\lan~\ran_B$ means some averaging over
(nonperturbative) background fields $B_\mu$. The precise form of
this averaging is not needed for our purpose. 

Integration over the ghost and gluon degrees of freedom in Eq. (\ref{25a})
yields the same answer as Eq. (\ref{4.14}), but where now all fields are
subject to the periodic boundary conditions (\ref{24a}).
$$
Z(B)=N'(\det W(B))_{reg}^{-1/2}[\det(-D_\mu(B)D_\mu(B+a))]_{a=
\frac{\delta}{\delta J}}
$$
\be 
\times \left \{1+\sum^\infty_{l=1} \frac{S_{int}
\left(a= \frac{\delta}{\delta J}\right)^l}{l!} 
\right\}\exp \left(
-\frac12JGJ\right)_{J_\mu=D_\mu(B)F_{\mu\nu}(B)}.
 \label{26a}
\ee 
We can consider strong background fields, so that $gB_\mu$ is
large (as compared to $\Lambda^2_{QCD}$), while $\alpha_s=
g^2/4\pi$ in that strong background is small at all distances.
Moreover, it was shown that $\alpha_s$ is frozen at large 
distances\cite{17}.
In this case Eq. (\ref{26a}) is a perturbative sum in powers of $g^n$,
arising from the expansion in $(ga_\mu)^n$.

In what follows we shall discuss the Feynman  graphs for the free
energy $F(T)$, connected to $Z(B)$ via 
\be 
F(T)=-T\ln \lan Z(B)\ran_B. 
\label{27a} 
\ee
 As will be seen, the lowest order
graphs  already contain a nontrivial dynamical mechanism  for the
deconfinement transition, and those will be considered in the next
subsection.

\subsection{The lowest order gluon contribution}

To lowest order in $ga_\mu$ (keeping all dependence on
$gB_\mu$ explicit) we have
\be 
Z_0=e^{-F_0(T)/T}=N'\lan \exp
(-F_0(B)/T)\ran_B, 
\label{28a} 
\ee
where using Eq. (\ref{26a}) $F_0(B)$ can be written as
\begin{eqnarray}
\frac{1}{T} F_0(B) &=&\frac12\ln \det G^{-1} -\ln \det (-D^2(B))=
\nonumber
\\
&=&Sp\left\{ -\frac12 \int^\infty_0 \xi (t) \frac{dt}{t}
e^{-tG^{-1}} + \int^\infty_0 \xi (t) \frac{dt}{t}
e^{tD^2(B)}\right\}. 
\label{29a} 
\end{eqnarray}

In Eq. (\ref{29a}) $Sp$ implies summing over all variables (Lorentz and
color indices and coordinates) and $\xi(t)$ is a regularization factor 
as in Eq. (\ref{8.4}).  
Graphically, the first term on the r.h.s. of Eq. (\ref{29a})
 is a gluon loop in the background field, while the second term is
a ghost loop.

Let us turn now to  the averaging procedure in Eq. (\ref{28a}). With
the notation $\varphi=-F_0(B)/T$, we can exploit in Eq. (\ref{28a})
the cluster expansion\cite{14}
 \begin{eqnarray} 
\lan \exp \varphi\ran_B&&=\exp
\left (\sum^\infty_{n=1} \lan \lan
\varphi^n\ran\ran\frac{1}{n!}\right) 
\nonumber
\\
&&=\exp \{ \lan
\varphi\ran_B+\frac12[ \lan \varphi\ran^2_B- \lan
\varphi^2\ran_B]+O (\varphi^3)\}. 
\label{30a}
\end{eqnarray} 
To get a closer
look at $\lan \varphi\ran_B$ we first should discuss the thermal
propagators of the gluon and ghost in the background field. We start
with the thermal ghost propagator and write the FSR for it\cite{11} 
\be 
(-D^2)^{-1}_{xy}= \lan x|\int^\infty_0
dte^{tD^2(B)}|y\ran= \int^\infty_0 dt(Dz)^w_{xy} e^{-K}\hat
\Phi(x,y). \label{31a}
\ee 
Here $\hat \Phi$ is the parallel
transporter in the adjoint representation along the trajectory of
the ghost: 
\be 
\hat \Phi(x,y) =P\exp (ig \int\hat B_\mu(z)
dz_\mu)
\label{32a}
\ee
and $(Dz)^w_{xy}$ is a path integration with boundary conditions
imbedded (denoted by the subscript $(xy)$) and with all
possible windings in the Euclidean temporal direction (denoted by the
superscript $w$).
We can write it explicitly as 
\begin{eqnarray} 
(Dz)^w_{xy}&&= \lim_{N\to
\infty}\prod^N_{m=1} \frac{d^4\xi(m)}{(4\pi\varepsilon)^2}
\nonumber
\\
&& \sum_{n=0,\pm,...} \frac{d^4p}{(2\pi)^4}\exp \left[ ip\left(
\sum^N_{m=1} \zeta(m)-(x-y)-n\beta \delta_{\mu 4}\right)\right].
\label{33a}
\end{eqnarray}
Here, $\zeta(k)=z(k)-z(k-1),~~ N\varepsilon=t$.
We can readily verify that in the free case, $ \hat B_\mu=0$, Eq. (\ref{31a})
reduces to the well-known form  of  the free propagator
\begin{eqnarray}
(-\partial^2)^{-1}_{xy}=&&\int^\infty_0 dt\exp\left
[-\sum^N_1\frac{\zeta^2(m)}{4\varepsilon}\right ] \prod_m
\overline{d\zeta(m)} \sum_n\frac{d^4p}{(2\pi)^4}
\nonumber
\\
 &&\times \exp \left [ ip \left ( \sum \zeta(m) -(x-y)
 -n\beta\delta_{\mu4}\right)\right] 
\nonumber
\\
 =&&\sum_n\int^\infty_0\exp \left
 [-p^2t-ip(x-z)-ip_4n\beta\right] dt \frac{d^4p}{(2\pi)^4}
 \label{34a}
 \end{eqnarray}
 with
 $$
 \overline{d\zeta(m)}\equiv \frac{d\zeta(m)}{(4\pi\varepsilon)^2}.
   $$
   Using the Poisson summation formula
   \be
   \frac{1}{2\pi}\sum_{n=0,\pm1, \pm2...} \exp (ip_4 n\beta) =
   \sum_{k=0,\pm1,...} \delta(p_4\beta-2\pi k)
   \label{35a}
   \ee
   we finally obtain the  standard  form
   \be
   (-\partial^2)^{-1}_{xy}=\sum_{k=0,\pm1,...}
   \int\frac{Td^3p}{(2\pi)^3}\frac{\exp[-ip_i(x-y)_i-i2\pi
   kT(x_4-y_4)]}{ p^2_i+(2\pi kT)^2}.
   \label{36a}
   \ee

   Note that, as expected, the propagators  (\ref{31a}) and (\ref{36a})
 correspond to a sum of ghost paths with all possible windings
 around the torus. The momentum integration in Eq. (\ref{33a}) asserts
 that the sum of all infinitesimal "walks" $\zeta(m)$ should be equal
 to the distance $(x-y)$ modulo $N$ windings in the
 compactified fourth coordinate.
 For the gluon propagator in the background gauge we
 obtain similarly to Eq. (\ref{31a})
 \be
 G_{xy}=\int^\infty_0 dt (Dz)^w_{xy} e^{-K}\hat
 \Phi_F(x,y),
 \label{37a}
 \ee
 where
 \be   
\hat \Phi_F(x,y)= P_F P\exp \left( -2ig \int^t_0 \hat
 F(z(\tau))d\tau\right) \exp \left( ig \int^x_y \hat
 B_\mu dz_\mu\right).
 \label{38a}
 \ee
The operators $P_F P$ are used to order insertions
 of $\hat F$ on the trajectory of the gluon.

 Now we come back to the first term in Eq. (\ref{30a}),
 $\lan\varphi\ran_B$, which can be representated with
 the help of Eqs. (\ref{31a})  and (\ref{37a}) as
 \be
 \lan \varphi\ran_B=\int\frac{dt}{t} \zeta(t) d^4
 x(Dz)^w_{xx} e^{-K}\left [ \frac12 tr \lan \hat
 \Phi_F(x,x)\ran_B-\lan tr \hat \Phi (x,x)\ran_B\right],
 \label{39a}
 \ee
 where $tr$ implies summation over Lorentz and
 color indices.
We can easily show\cite{11} that Eq. (\ref{39a}) yields for
 $B_\mu=0$ the usual result for  the free gluon gas:
 \be
 F_0(B=0)=-T\varphi(B=0)=-(N^2_c-1)
 V_3\frac{T^4\pi^2}{45}.
 \label{40a}
 \ee

 \subsection{The lowest order quark contribution}

 Integrating over the quark fields in Eq. (\ref{25a})
 leads to the following additional factor in Eq. (\ref{26a})
 \be
 \det(m+\hat D(B+a))=[\det (m^2-\hat D^2(B+a))]^{1/2}.
 \label{41a}
\ee
%where we have used the symmetry  property of eigenvalues of $\hat D$. 
In the lowest approximation, we may omit $a_\mu$ in Eq. (\ref{41a}).
As a result we get a contribution from the quark fields to the free energy
 \be
 \frac{1}{T} F^q_0(B)=-\frac12\ln\det (m^2-\hat D^2(B))=-\frac12
 Sp\int^\infty_0\xi(t)\frac{dt}{t} e^{-tm^2+t\hat D^2(B)},
 \label{42a}
 \ee
 where $Sp$ has the same meaning as in Eq. (\ref{29a}) and
 \begin{eqnarray}
 \hat D^2=(D_\mu\gamma_\mu)^2=&&D^2_\mu(B)-gF_{\mu\nu}\sigma_{\mu\nu}
 \equiv D^2-g\sigma F;
\nonumber
\\
&&\sigma_{\mu\nu} =+\frac{i}{4}
 (\gamma_\mu\gamma_\nu-\gamma_\nu\gamma_\mu).
 \label{43a}
 \end{eqnarray}
 Our aim now is to exploit the FSR to represent Eq. (\ref{42a}) in a form
 of the path integral, as was done for gluons in Eq. (\ref{31a}). The
 equivalent form for quarks must implement the antisymmetric boundary
 conditions pertinent to fermions. We find
 \be
 \frac{1}{T} F^q_0(B)=-\frac12 tr \int^\infty_0\xi(t) \frac{dt}{t}
 d^4x
 \overline{(Dz)}^w_{xx}e^{-K-tm^2}W_\sigma(C_n),
 \label{44a}
  \ee
  where
$$
  W_\sigma(C_n)=P_FP_A\exp\left (ig \int_{C_n} A_\mu dz_\mu\right)
  \exp g\left (\sigma F\right), 
$$ 
and
 \begin{eqnarray}
 \overline{(Dz)}^w_{xy}&&=\prod^N_{m=1}\frac{d^4\zeta(m)}{(4\pi
 \varepsilon)^2}
\nonumber
\\
&& \!\!\!\sum_{n=0,\pm1,\pm2,...}\!\!\!
 (-1)^n\frac{d^4p}{(2\pi)^4}
 \exp \left [ip\left(\sum^N_{m=1}
 \zeta(m)-(x-y)-n\beta
 \delta_{\mu 4}\right)\right].
 \label{45a}
 \end{eqnarray}

It can readily be checked that in the case $B_\mu=0$
 the well known expression for the free quark gas is recovered, i.e.
 \be
 F^q_0({\rm free~quark}) =-\frac{7\pi^2}{180} N_cV_3T^4\cdot n_f,
 \label{46a}
 \ee
 where $n_f$ is the number of flavors. The derivation of Eq. (\ref{46a})
 starting from the path-integral form (\ref{44a}) is done similarly to
 the gluon case given in the Appendix of the last reference in Ref.
\cite{11}.

 The loop $C_n$ in Eq. (\ref{44a})  corresponds to $n$ windings in the
 fourth direction. Above the deconfinement transition temperature
 $T_c$ one sees in Eq. (\ref{44a}) the appearance of the  factor
 \be
 \Omega=P \exp \left [ig \int^\beta_0 B_4(z) dz_4\right].
 \label{47a}
 \ee
For the constant field $B_4$ and
 $B_i=0,i=1,2,3, $ we obtain
\be 
\lan F\ran=-\frac{V_3}{\pi^2}
 tr_c\sum^\infty_{n=1} \frac{\Omega^n+\Omega^{-n}}{n^4} (-1)^{n+1}.
 \label{48a}
 \ee
 This result coincides with the one obtained in the literature\cite{42}.

\section{Discussion and conclusions}

Three basic approaches to QCD which
are largely used till now are: i) lattice simulations ii) standard 
perturbation theory, and iii) OPE and  QCD sum rules\cite{32}. 
The two latter methods are analytic and have 
given enormous amount of theoretical information
 about the high-energy domain, where perturbative methods are applicable, 
and about nonperturbative effects both in the high and low energy regions.

These methods have their own limitations. In particular, the standard perturbation 
theory is plagued by the Landau ghost pole and IR renormalons
and slow convergence, which necessitates  the introduction of methods,
 where summation of perturbative  subseries can be done automatically and the
Landau ghost pole is absent. 
The QCD sum rules are limited by the use of only a few OPE terms, while the OPE
 series is known to be badly convergent and at best asymptotic.

One of the great challenges of QCD is to have a tractable analytic treatment
of it. In particular, the improvement of the standard perturbation theory 
and the search for a systematic approach to nonperturbative phenomena are 
important objectives.
The methods presented here in the present paper, commonly entitled 
The Fock-Feynman-Schwinger Representation, are  meant to exactly do this.
The main advantage of the FSR is that it allows to treat
both perturbative and nonperturbative configurations of the gluonic fields.

In case of purely perturbative fields the FSR yields a simple 
method of summation and exponentiation of perturbative  diagrams\cite{15}.
Nonperturbative fields are  introduced in the FSR naturally
 via the Field Correlator Method\cite{38,55}. A recent discovery on
 the lattice of the Gaussian correlator dominance (Gaussian Stochastic Model) 
(see Ref. \cite{56} for discussion and further references.) makes 
this method accurate (up to a few percent).
There is another very important result of taking into account 
nonperturbative fields
in the QCD vacuum: this fact allows to develop perturbation theory in the 
nonperturbative background -- which is realistic unlike the standard perturbation 
theory. It contains no Landau ghost poles and IR renormalons\cite{17}.

As two applications of the FSR we have considered the problem of collinear
singularities and finite temperature QCD.
As an important special feature we should stress the absence in the 
background perturbation theory of all IR and collinear singularities 
pertinent to standard perturbation theory.
This feature discussed here, opens new perspectives to
 the application of FSR to high-energy QCD processes. Although not discussed
here, the FSR can readily be extended
to treat deep inelastic scattering,
Drell-Yan and other processes, including the fundamental problem
 of the connection between constituent quark-gluon model and parton model.

Moreover, as shown here the FSR can be used to describe QCD 
at nonzero temperature at above and around phase transition point. 
As was shown before\cite{11,12,13} the vacuum is predominantly 
magnetic and nonperturbative above $T_c$. Therefore methods based on FSR are 
working well in this region. 
In conclusion, since the FSR is replacing field degrees of freedom by 
corresponding quantum mechanical ones it has the advantage, that the results
can often be interpreted in a simple and transparant way.
It has been applied with success to both Abelian and non-Abelian situations.
We have found that the FSR is a powerful approach for studying problems in QCD.

\section{Acknowledgements}

This work was started while one of the authors (Yu.S)
 was a guest of the Institute for Theoretical Physics of Utrecht University.
The kind hospitality of the Institute and all persons involved and useful
discussions with N. van Kampen, Th. Ruijgrok  and G. 't Hooft are gratefully
acknowledged. The authors have been partially supported by the grant
INTAS 00-110. One of the authors (J.T) would like to thank the TQHN group at the
University of Maryland and the theory group at TJNAF for their kind hospitality.
Yu. S. was partially supported by the RFFI grants 00-02-17836 and 00-15-96786
and also by the DOE contract DE-AC05-84ER40150 under which SURA operates the TJNAF.


\begin{thebibliography}{99}
\bibitem{1}
R. P. Feynman, Phys. Rev. {\bf 80} (1950), 440; ibid {\bf 84} (1951), 108.
\bibitem{2}
V. A. Fock, Izvestya Akad. Nauk USSR, OMEN, 1937, p.557; J. Schwinger,
Phys. Rev. {\bf 82} (1951), 664.
\bibitem{3} G. A. Milkhin and E.S. Fradkin, ZhETF {\bf 45} (1963), 1926; 
E. S. Fradkin, Trudy FIAN, {\bf 29} (1965), 7.
\bibitem{4} M. B. Halpern, A. Jevicki and P. Senjovic, Phys. Rev.  {\bf D16} (1977), 2474; 
R. Brandt et al., Phys. Rev. {\bf D19} (1979), 1153; 
S. Samuel, Nucl. Phys. {\bf B149} (1979), 517;
     J. Ishida and A. Hosoya, Progr. Theor. Phys. {\bf 62} (1979), 544.
\bibitem{5} Yu. A. Simonov, Nucl. Phys. {\bf B307} (1988), 512.
\bibitem{6} Yu. A. Simonov, Yad. Fiz.{\bf 54} (1991), 192.
\bibitem{7} A. I. Karanikas and C. N. Ktorides, Phys. Lett.  {\bf B275} (1992), 403; 
Phys. Rev. {\bf D52} (1995), 58883;
A. I. Karanikas, C. N. Ktorides and N. G. Stefanis, Phys.  Rev. {\bf D52} (1995), 5898.
\bibitem{8} Z. Bern and D. A. Kosower, Phys. Rev. Lett. {\bf 66}
      (1991), 1669; Nucl. Phys. {\bf B379} (1992), 451;
      Z. Bern and D. C. Dunbar, Nucl. Phys. {\bf B379} (1992), 562.
\bibitem{9} M. J. Strassler, Nucl. Phys. {\bf B385} (1992), 145.
\bibitem{10} M. Reuter, M. G. Schmidt and C. Schubert, Ann.Phys. (NY) {\bf 259} (1997), 313;
H.-T. Sato and M. G. Schmidt,{\bf B560} (1999), 551.
\bibitem{11} Yu. A. Simonov, JETP Lett {\bf 54} (1991), 249; ibid. {\bf 55} (1992), 627;
Phys. At. Nucl. {\bf 58} (1995), 309.
\bibitem{12} Yu. A. Simonov, in "Varenna 1995, Selected
topics in nonperturbative QCD", p. 319, 1995.
\bibitem{13}
H. G. Dosch, H.-J. Pirner and Yu. A. Simonov, Phys. Lett. {\bf B349}
      (1995), 335; E. L. Gubankova and Yu. A. Simonov, Phys.
      Lett. {\bf B360} (1995), 93.\\
      N. O. Agasyan, D. Ebert and E. M. Ilgenfritz, Nucl.Phys.{\bf A637} (1998), 135.
      \bibitem{14} N. G. van  Kampen, Phys. Rep. {\bf 24 C} (1976), 171.
      \bibitem{15} Yu. A. Simonov, Phys. Lett. {\bf B464 } (1999), 265.
      \bibitem{16}
 B. S. De Witt, Phys. Rev. {\bf 162} (1967), 1195\\
 J. Honerkamp, Nucl. Phys.  {\bf B48} (1972), 269;\\
 G. 't Hooft Nucl. Phys.  {\bf B62} (1973), 444; in "Lectures
at Karpacz" Acta Univ. Wratislaviensis {\bf 368} (1976), 345;\\
L. F. Abbot, Nucl. Phys.  {\bf B185} (1981), 189.

\bibitem{17} Yu. A. Simonov, Phys. At Nucl. {\bf 58} (1995), 107; \\
JETP Lett.  {\bf 75} (1993), 525.
Yu. A. Simonov, in: "Lecture Notes in Physics" 
(H. Latal and W. Schweiger, Eds.), Vol. 479, p. 138, Springer, 1996.
\bibitem{18} Yu. A. Simonov and J. A. Tjon, Ann Phys. {\bf 228} (1993), 1.
\bibitem{19} J. A. Tjon, Quark Confinement and hadron spectrum III, ed.
N. Isgur, World Scientific, Singapore, p 113, 2000.
\bibitem{20} T. Nieuwenhuis, J. A. Tjon and Yu. A. Simonov, Few Body
Systems, Suppl. {\bf 7} (1994), 286.
\bibitem{21} T.~Nieuwenhuis, PhD-thesis, University of Utrecht (1995), unpublished.
\bibitem{22} T.~Nieuwenhuis and J. A.~Tjon, Phys.~Lett.~{\bf B355}
(1995), 283.
\bibitem{23} T.~Nieuwenhuis and J. A.~Tjon, Phys. Rev. Lett. {\bf 77}
(1996), 814.
\bibitem{24} \c{C}. \c{S}avkl{\i}, F.~Gross, and J. A.~Tjon, Phys. Rev.
{\bf D62} (2000), 116006. 
\bibitem{25} \c{C}. \c{S}avkl{\i}, F.~Gross, and J. A.~Tjon, Phys. 
Lett. {\bf B62} (2002), 116. 

\bibitem{26} Yu. A. Simonov, J. A. Tjon, hep-ph/0201005.
\bibitem{27}
V. S. Dotsenko, S. N. Vergeles, Nucl. Phys.  {\bf B169} (1980), 527.

\bibitem{34} Yu. A. Simonov, Phys. Rep. {\bf 320 C} (1999), 265; JETP
Lett. {\bf 69} (1999), 505.
\bibitem{38}
  H. G. Dosch, Phys. Lett. {\bf B190} (1987), 177;\\
  H. G. Dosch and Yu. A. Simonov, Phys. Lett. {\bf B205} (1988), 339;\\
   Yu. A. Simonov, Nucl.  Phys.  {\bf B307} (1988), 512.

\bibitem{35}
 Yu. A. Simonov, S. Titard and F. J. Yndurain, Phys. Lett. {\bf B359 } (1995), 435.
\bibitem{37}
Yu. A. Simonov, Nucl. Phys.  {\bf B324} (1989), 67;
 H. G. Dosch and M. Schiestl, Phys. Lett. {\bf B209} (1988), 85.

 \bibitem{39} M. Shifman, A. I. Vainshtein and V. I. Zakharov, Nucl. Phys.
 {\bf B147} (1979), 385.

 \bibitem{40}
A. Di Giacomo and H. Panagopoulos, Phys. Lett. {\bf B285} (19992), 133;
 A. Di Giacomo, E. Meggiolaro and H. Panagopoulos, Nucl. Phys.
{\bf B483} (1997), 371.

\bibitem{36}
M. B. Voloshin, Nucl. Phys. {\bf B154} (1979), 365, Sov. J. Nucl.
Phys. {\bf 36} (1982), 143;\\
H. Leutwyler, Phys. Lett. {\bf B98} (1981), 447.

\bibitem{41}
A. M. Badallian and V. P. Yurov, Yad. Fiz. {\bf 51} (1990), 1368; Phys.
Rev. {\bf D42} (1990), 3138.

\bibitem{42} G. J. Gross, R. D. Pisarski and L. G. Yaffe, Rev. Mod. Phys.
{\bf 53} (1981), 43.

\bibitem{28} A. M. Badalian and V. I. Morgunov, Phys. Rev. {\bf D60}
(1999), 116008;\\
A. M. Badalian and B. L. G. Bakker, Phys. Rev. {\bf D62} (2000), 94031.

\bibitem{29}
A. Yu. Dubin, A. B. Kaidalov, and Yu. A. Simonov, Phys. Lett. {\bf B343}
(1995), 310; Phys. Atom. Nucl.  {\bf 58} (1995), 300.

\bibitem{30}
V. L. Morgunov, V. I. Shevchenko and Yu. A. Simonov, Phys. At. Nucl. {\bf
61} (1998), 664; Phys. Lett. {\bf B343} (1995), 310.

\bibitem{31} Yu. A. Simonov,  Phys. At. Nucl. {\bf 60} (1997), 2069,
hep-ph/9704301;\\
Yu.A.~Simonov and J.A.~Tjon,  Phys.\ Rev.  {\bf
D62} (2000), 014501; ibid.  {\bf D62} (2000), 094511.
\bibitem{32} Yu. A. Simonov,  Phys. At. Nucl. {\bf 62} (1999), 1932,
hep-ph/9912383.
\bibitem{33}
A. Yu. Dubin, A. B. Kaidalov, and Yu. A. Simonov, Phys. Lett. {\bf B323}
(1994), 41.
\bibitem{43}
A. I. Akhiezer and V. B. Berestetskii, Quantum Electrodynamics, 
New York, Interscience Publishers, 1965.
\bibitem{44}
F. J. Yndurain, The theory of Quark and Gluon Interactions, 3rd
edition, Springer, Chapter 5, p. 189, 1998.
\bibitem{45}
T. Kinoshita, J.Math. Phys. {\bf 3} (1962), 650;\\
T. D. Lee and M. Nauenberg, Phys. Rev. {\bf 133} (1964), B1549.
\bibitem{56}
Yu. A. Simonov, JETP Lett. {\bf 71} (2000) 127;\\ 
V. I. Shevchenko and Yu. A. Simonov Phys. Rev. Lett. {\bf 85} (2000), 1811.
\bibitem{46}
A. M. Badalian and Yu. A. Simonov, Yad. Fiz. {\bf 60} (1997), 714.
\bibitem{50}
G. Sterman and S. Weinberg, Phys. Rev. Lett. {\bf 39} (1977), 1436.

\bibitem{53}
 A. B. Kaidalov and   Yu. A. Simonov, Phys. Lett. {\bf B477} (2000), 163; 
Phys. At. Nucl. {\bf 63} (2000), 1428.

\bibitem{47} S. G. Gorishny, A. L. Kataev and S. A. Larin, Phys. Lett.
{\bf B259} (1991), 144;\\
M. A. Samuel and L. P. Surguladze, Phys. Rev. Lett. {\bf 66} (1991), 560.

\bibitem{48} G. t' Hooft, Nucl. Phys. {\bf B72} (1974), 461.

\bibitem{49} P. Cea, G. Nardulli and G. Preparata, Z. Phys. {\bf C16}
(1982) 135;\\ Phys. Lett. {\bf B115} (1982), 310.

\bibitem{51} R. K. Ellis, QCD at TASI '94, Fermilab-Conf-94/410-T.

\bibitem{52}
S. Bethke et al. Phys. Lett. {\bf B213} (1988), 235.

\bibitem{54} OPAL collaboration: Zeit. Phys. {\bf C49} (1991), 375.

\bibitem{55} A. Di Giacomo, H. G. Dosch,V. I. Shevchenko and Yu. A. Simonov,
Phys. Rep. (in press), hep-ph/0007223.

\end{thebibliography}
\end{document}